\documentclass[a4paper,11pt]{article}
\pdfoutput=1
\usepackage{jheppub}

\usepackage[MeX,plmath]{polski}
\usepackage[utf8]{inputenc}
\usepackage[polish,american]{babel}
\usepackage{braket}

\usepackage{tikz}
\usepackage{multirow}
\usepackage{array}
\newcolumntype{L}[1]{>{\raggedright\let\newline\\\arraybackslash\hspace{0pt}}m{#1}}
\newcolumntype{C}[1]{>{\centering\let\newline\\\arraybackslash\hspace{0pt}}m{#1}}
\newcolumntype{R}[1]{>{\raggedleft\let\newline\\\arraybackslash\hspace{0pt}}m{#1}}

\usepackage{lineno}

\usepackage{arydshln}

\newcommand{\alktev}{A_L = (3.322\pm 0.058_{stat} \pm 0.047_{syst}) \times 10^{-3}}
\newcommand{\askloeold}{A_S = (1.5 \pm 9.6_{stat} \pm 2.9_{syst}) \times 10^{-3}}

\newcommand{\ksAs}{-4.9}
\newcommand{\ksAsErrStat}{5.7}
\newcommand{\ksAsErrSyst}{2.6}
\newcommand{\asresult}{A_S = (\ksAs \pm \ksAsErrStat_{stat} \pm \ksAsErrSyst_{syst}) \times 10^{-3}}
\newcommand{\asresultcombined}{A_S = (-3.8 \pm 5.0_{stat} \pm 2.6_{syst}) \times 10^{-3}}

\newcommand{\cutClusterEnergy}{ 100~ \mbox{MeV}}
\newcommand{\cutBeta}{ 0.18 < \beta^{*} < 0.27  }

\newcommand{\cutRho}{ \rho_{vtx} < 15 \mbox{ cm}}
\newcommand{\cutZ}{ |z_{vtx}| < 10 \mbox{ cm}}

\newcommand{\ksParEn}{36874}
\newcommand{\ksParEp}{34579}

\newcommand{\ksParErrEn}{255}
\newcommand{\ksParErrEp}{251}

\newcommand{\ksEffiEp}{7.39}
\newcommand{\ksEffiEn}{7.81}
\newcommand{\ksEffiErrEp}{0.03}
\newcommand{\ksEffiErrEn}{0.03}
\newcommand{\ksEffiEpFilfo}{99.80}
\newcommand{\ksEffiEpTag}{36.54}
\newcommand{\ksEffiEpDC}{75.60}
\newcommand{\ksEffiEpTCA}{42.22}
\newcommand{\ksEffiEpTOF}{64.03}
\newcommand{\ksEffiEpOther}{99.16}
\newcommand{\ksEffiEnFilfo}{99.80}
\newcommand{\ksEffiEnTag}{36.67}
\newcommand{\ksEffiEnDC}{75.62}
\newcommand{\ksEffiEnTCA}{41.85}
\newcommand{\ksEffiEnTOF}{67.96}
\newcommand{\ksEffiEnOther}{99.17}
\newcommand{\ksEffiErrEpFilfo}{0.02}
\newcommand{\ksEffiErrEpTag}{0.05}
\newcommand{\ksEffiErrEpDC}{0.07}
\newcommand{\ksEffiErrEpTCA}{0.08}
\newcommand{\ksEffiErrEpTOF}{0.19}
\newcommand{\ksEffiErrEpOther}{0.03}
\newcommand{\ksEffiErrEnFilfo}{0.02}
\newcommand{\ksEffiErrEnTag}{0.05}
\newcommand{\ksEffiErrEnDC}{0.07}
\newcommand{\ksEffiErrEnTCA}{0.08}
\newcommand{\ksEffiErrEnTOF}{0.18}
\newcommand{\ksEffiErrEnOther}{0.02}

\newcommand{\ksChiAll}{118/109 }

\newcommand{\Messina}{a}
\newcommand{\Frascati}{b}
\newcommand{\Warsaw}{c}
\newcommand{\INFNRomaThree}{d}
\newcommand{\RomaThree}{e} 
\newcommand{\Uppsala}{f}
\newcommand{\Cracow}{g}
\newcommand{\RomaOne}{h}
\newcommand{\INFNRomaOne}{i}
\newcommand{\Marconi}{j}
\newcommand{\INFNRomaTwo}{k}
\newcommand{\RomaTwo}{l}
\newcommand{\GSSI}{m}
\newcommand{\ENEACasaccia}{n}
\newcommand{\BINP}{o}
\newcommand{\MessinaTwo}{p}
\newcommand{\Novosibirsk}{r}
\newcommand{\INFNCatania}{s}
\newcommand{\Energetica}{t}
\newcommand{\INFNBari}{u}
\newcommand{\Calabria}{v}
\newcommand{\INFNCalabria}{w}
\newcommand{\IFIN}{x}
\newcommand{\INFNPisa}{y}

\newcommand{\affuni}[2]{Dipartimento di Fisica dell'Universit\`a #1, #2, Italy.}
\newcommand{\affinfn}[2]{INFN Sezione di #1, #2, Italy.}

\affiliation[\Messina]{Dipartimento di Scienze Matematiche e Informatiche, Scienze Fisiche e Scienze della Terra dell'Universit\`a di Messina, Messina, Italy.}
\affiliation[\Frascati]{Laboratori Nazionali di Frascati dell'INFN, Frascati, Italy.}
\affiliation[\Warsaw]{National Centre for Nuclear Research, Warsaw, Poland.}
\affiliation[\INFNRomaThree]{\affinfn{Roma Tre}{Roma}}
\affiliation[\RomaThree]{Dipartimento di Matematica e Fisica dell'Universit\`a  ``Roma Tre'', Roma, Italy.}
\affiliation[\Uppsala]{Department of Physics and Astronomy, Uppsala University, Uppsala, Sweden.}
\affiliation[\Cracow]{Institute of Physics, Jagiellonian University, Cracow, Poland.}
\affiliation[\RomaOne]{\affuni{``Sapienza''}{Roma}}
\affiliation[\INFNRomaOne]{\affinfn{Roma}{Roma}}
\affiliation[\Marconi]{Dipartimento di Scienze e Tecnologie applicate, Universit\`a ``Guglielmo Marconi", Roma, Italy.}
\affiliation[\INFNRomaTwo]{\affinfn{Roma Tor Vergata}{Roma}}
\affiliation[\RomaTwo]{\affuni{``Tor Vergata''}{Roma}}
\affiliation[\GSSI]{Gran Sasso Science Institute, L'Aquila, Italy.}
\affiliation[\ENEACasaccia]{ENEA UTTMAT-IRR, Casaccia R.C., Roma, Italy}
\affiliation[\BINP]{Budker Institute of Nuclear Physics, Novosibirsk, Russia.}
\affiliation[\MessinaTwo]{Dipartimento di Scienze Chimiche, Biologiche, Farmaceutiche ed Ambientali dell'Universit\`a di Messina, Messina, Italy.}
\affiliation[\Novosibirsk]{Novosibirsk State University, Novosibirsk, Russia.}
\affiliation[\INFNCatania]{\affinfn{Catania}{Catania}}
\affiliation[\Energetica]{Dipartimento di Scienze di Base ed Applicate per l'Ingegneria dell'Universit\`a ``Sapienza'', Roma, Italy.}
\affiliation[\INFNBari]{\affinfn{Bari}{Bari}}
\affiliation[\Calabria]{\affuni{della Calabria}{Rende}}
\affiliation[\INFNCalabria]{INFN Gruppo collegato di Cosenza, Rende, Italy.}
\affiliation[\IFIN]{Horia Hulubei National Institute of Physics and Nuclear Engineering, M\v{a}gurele, Romania}
\affiliation[\INFNPisa]{\affinfn{Pisa}{Pisa}}

\collaboration{The KLOE-2 Collaboration}
\author[\Messina,\Frascati]{A.~Anastasi}
\author[\Frascati]{D.~Babusci}
\author[\Frascati,\Warsaw]{M.~Ber\l{}owski}
\author[\Frascati]{C.~Bloise}
\author[\Frascati]{F.~Bossi}
\author[\INFNRomaThree]{P.~Branchini}
\author[\RomaThree,\INFNRomaThree]{A.~Budano}
\author[\Uppsala]{B.~Cao}
\author[\Frascati]{G.~Capon}
\author[\RomaThree,\INFNRomaThree]{F.~Ceradini}
\author[\Frascati]{P.~Ciambrone}
\author[\Frascati]{F.~Curciarello}
\author[\Cracow]{E.~Czerwi\'nski}
\author[\RomaOne,\INFNRomaOne]{G.~D'Agostini}
\author[\Frascati]{E.~Dan\`e}
\author[\INFNRomaTwo]{V.~De~Leo}
\author[\Frascati]{E.~De~Lucia}
\author[\Frascati]{A.~De~Santis}
\author[\Frascati]{P.~De~Simone}
\author[\RomaThree,\INFNRomaThree]{A.~Di~Cicco}
\author[\RomaOne,\INFNRomaOne]{A.~Di~Domenico}
\author[\Frascati]{D.~Domenici}
\author[\Frascati]{A.~D'Uffizi}
\author[\RomaTwo,\INFNRomaTwo]{A.~Fantini}
\author[\GSSI]{G.~Fantini}
\author[\Frascati]{P.~Fermani}
\author[\ENEACasaccia,\INFNRomaOne]{S.~Fiore}
\author[\Cracow]{A.~Gajos}
\author[\RomaOne,\INFNRomaOne]{P.~Gauzzi}
\author[\Frascati]{S.~Giovannella}
\author[\INFNRomaThree]{E.~Graziani}
\author[\BINP,\Novosibirsk]{V.~L.~Ivanov}
\author[\Uppsala]{T.~Johansson}
\author[\Frascati]{X.~Kang}
\author[\Cracow]{D.~Kisielewska-Kami\'nska}
\author[\BINP,\Novosibirsk]{E.~A.~Kozyrev}
\author[\Warsaw]{W.~Krzemie\'n}
\author[\Uppsala]{A.~Kup\'s\'c}
\author[\RomaThree,\INFNRomaThree]{S.~Loffredo}
\author[\BINP,\Novosibirsk]{P.~A.~Lukin}
\author[\MessinaTwo,\INFNCatania]{G.~Mandaglio}
\author[\Frascati,\Marconi]{M.~Martini}
\author[\RomaTwo,\INFNRomaTwo]{R.~Messi}
\author[\Frascati]{S.~Miscetti}
\author[\INFNRomaTwo]{D.~Moricciani}
\author[\Cracow]{P.~Moskal}
\author[\INFNRomaThree]{A.~Passeri}
\author[\Energetica,\INFNRomaOne]{V.~Patera}
\author[\Frascati]{E.~Perez~del~Rio}
\author[\INFNRomaTwo]{N.~Raha}
\author[\Frascati]{P.~Santangelo}
\author[\Calabria,\INFNCalabria]{M.~Schioppa}
\author[\RomaThree,\INFNRomaThree]{A.~Selce}
\author[\Cracow]{M.~Silarski}
\author[\Frascati,\IFIN]{F.~Sirghi}
\author[\BINP,\Novosibirsk]{E.~P.~Solodov}
\author[\INFNRomaThree]{L.~Tortora}
\author[\INFNPisa]{G.~Venanzoni}
\author[\Warsaw]{W.~Wi\'slicki}
\author[\Uppsala]{M.~Wolke}

\abstract{
	Using 1.63~fb$^{-1}$ of integrated luminosity  collected by the KLOE 
    experiment
    about 
	$7\times 10^4$  
    $K_S\rightarrow \pi^{\pm}e^{\mp}\nu$ 
    decays have been reconstructed.
	The measured value of the charge asymmetry for this decay is $\asresult$, which is almost twice more precise than the
	previous KLOE result. The combination of these two measurements gives $\asresultcombined$ 
    and, together with the asymmetry of the $K_L$ semileptonic decay, 
    provides significant tests of the $CPT$ symmetry.
The obtained results are in agreement with $CPT$ invariance.
    }

\toccontinuoustrue

\begin{document}

\title{Measurement of the charge asymmetry for the $K_S \rightarrow \pi e \nu$ decay 
and test of CPT symmetry with the KLOE detector}

\maketitle

\section{Introduction}
Semileptonic decays have been 
of fundamental importance in establishing several properties of 
the neutral kaon system, and of the Standard Model in general, including 
the $\Delta S = \Delta Q$ rule~\cite{dsdq},
$CP$ violation 
\cite{cpviolsemilep},
and the unitarity of the quark mixing matrix ~\cite{kloevus,flavianet}.
\par
The  asymmetries which can be constructed from the decay rates 
into the two $CP$ conjugated semileptonic final states,
$\pi^{-} e^{+} \nu$ and $\pi^{+} e^{-} \bar{\nu}$, constitute a
powerful probe 
 in the study of discrete symmetries \cite{sanda}.
%
In particular, the charge asymmetries for the physical states $K_S$ and $K_L$ defined as:
\begin{equation}
    \begin{aligned}
        A_{S,L}
        & = 
        \frac{\Gamma(K_{S,L} \rightarrow \pi^{-} e^{+} \nu) - \Gamma(K_{S,L}
            \rightarrow
            \pi^{+} e^{-}
        \bar{\nu})}{\Gamma(K_{S,L} \rightarrow \pi^{-} e^{+} \nu) + \Gamma(K_{S,L}
            \rightarrow
        \pi^{+} e^{-} \bar{\nu})}   
    \end{aligned}
\end{equation}
are sensitive to $CP$ violation effects. At first order in small parameters \cite{handbook_cp}:
\begin{equation}
    \begin{aligned}
        A_{S,L}   & = 2 \left[  Re\left( \epsilon_{K}\right) \pm Re \left(\delta_{K} \right) - Re (y) \pm Re( x_{-}) \right]       
    \end{aligned}
\end{equation}
with $Re\left(\epsilon_{K} \right) $ and $Re \left(\delta_{K} \right)$ implying $T$- and $CPT$-violation in the $K^0-\overline{K^0}$ mixing, respectively,
$Re( y)$ and $Re( x_-)$ implying $CPT$ violation in $\Delta S=\Delta Q$ and $\Delta S \neq \Delta Q$ decay amplitudes, 
respectively\footnote{More explicitly $y$ and $x_-$ are described in terms of the decay amplitudes
$ \mathcal{A_{\pm}} = A(K^0 \rightarrow e^{\pm} \pi^{\mp} \nu (\bar{\nu}))$ and 
$ \mathcal{\bar{A}_{\pm}} = A(\bar{K^0} \rightarrow e^{\pm} \pi^{\mp} \nu (\bar{\nu}))$ as:
\begin{align}
    \begin{aligned}
        y = \frac{ \mathcal{ \bar{A}_{-}^{*} } - \mathcal{A}_{+}  }{  \mathcal{ \bar{A}_{-}^{*} } +
        \mathcal{A}_{+} }, & &  
        x_{-}=  \frac{1}{2}  \left[ \frac{\mathcal{\bar{A}}_+}{\mathcal{A}_+}  - 
        \left( \frac{\mathcal{A}_-}{\mathcal{\bar{A}}_-} \right)^* \right]. 
    \end{aligned}
\end{align}
},
and all parameters implying $CP$ violation.
If $CPT$ symmetry holds then 
the two asymmetries are expected to be identical
$A_{S}\!=\!A_{L}\!=\!2\,\mathrm{Re}\,(\epsilon_K)\!\simeq\!3\!\times\!10^{-3}$
each accounting for the $CP$ impurity in the mixing in the corresponding physical state.
\par
The $CPT$ theorem ensures exact $CPT$ invariance for quantum field theories - like the Standard Model - formulated on flat space-time and 
assuming Lorentz invariance, locality, and hermiticity \cite{luders}.
$CPT$ violation effects might arise in a quantum gravity scenario \cite{mavroreview,liberatireview} and their observation
would constitute an unambiguous signal of processes beyond the Standard Model.
\par
In this context the measurement of the difference $A_{S}-A_{L}=4\,\left(\mathrm{Re}\,\delta_K+\mathrm{Re}\,x_{-}\right)$ is of particular importance
as a test of the $CPT$ symmetry. This observable is  
well constrained and can provide a test
based on the direct comparison of a transition probability with its $CPT$ conjugated transition - realised with entangled neutral kaon pairs -
which constitutes one of the most precise, robust and model independent tests of the $CPT$ symmetry
\cite{cpttrans}.
\par
The sum $A_{S}+A_{L}=4\left(\mathrm{Re}\,\epsilon_K-\mathrm{Re}\,y\right)$ can be used to extract the $CPT$-violating parameter 
$Re (y)$ once the measured value of $Re (\epsilon_K)$ is provided as input.
\par
The two combinations $A_{S}\pm A_L$ (dominated by the uncertainty on $A_S$) constitute also a fundamental ingredient 
for improving the semileptonic decay
contribution to the $CPT$ test obtained imposing the unitarity relationship,
originally derived by Bell and Steinberger \cite{BS},
and yielding the most stringent limits on $Im(\delta)$ and the mass difference $m(K^0)-m(\overline{K^0})$~\cite{Ambrosino:2006ek,Patrignani:2016xqp}. 
\par 
At present, the most precise measurement of $A_L$
has been performed by the KTeV collaboration: $\alktev$~\cite{AlaviHarati:2002bb}.
The measurement of its counterpart, $A_S$, 
requires a very pure $K_S$ beam which can only be realised exploiting 
the entangled neutral kaons pairs produced at a $\phi$-factory \cite{DiDomenico:2007zza}.
\\
The first measurement of $A_S$ has been performed by the KLOE collaboration 
using
410 pb$^{-1}$ of integrated luminosity 
  collected at DA$\Phi$NE \cite{dafne}, the $\phi$-factory of the INFN laboratories of Frascati: $\askloeold$~\cite{Ambrosino:2006si},
with an accuracy dominated by the statistical uncertainty.
The new measurement reported here is based on a four times larger data sample, 
corresponding to an integrated luminosity of $1.63$ fb$^{-1}$ collected in 2004-2005. The combination of the two results has
a precision
approaching the level of the  $CP$ violation effects expected for $K_S$ under the assumption of $CPT$ invariance.
New limits on $Re( y)$ and $Re( x_-)$ have been also derived. 

 \section{The KLOE detector}
 The KLOE detector operates at the DA$\Phi$NE electron-positron collider.
 The energy of the two colliding beams is set to the mass of the $\phi$ meson which 
 decays predominantly into a pair of charged or neutral kaons.
 Since the beams cross at an angle of $2 \times 12.5$~mrad the $\phi$-meson is produced with a small momentum~of $p_{\phi}  \approx 13$~MeV.

 The KLOE detector consists of two main components: the cylindrical drift chamber and the 
 electromagnetic calorimeter, both surrounding the beam pipe and immersed in a 0.52~T
axial magnetic field.
 The drift chamber (DC) is a $3.3\ \mbox{m}$ long cylinder with internal and external radii of
 $25\ \mbox{cm}$ and $2\ \mbox{m}$, respectively.
 The chamber structure is made of carbon-fiber epoxy composite
and the gas mixture used is $90\%$ helium, $10\%$ isobutane.
These features maximize transparency to photons and reduce  charged particle multiple scattering
and
$K_L \rightarrow K_S$
regeneration.
About $40\%$ of produced $K_L$ mesons decay inside the DC volume, while most of the 
surviving $K_L$'s interact
and are detected
in the electromagnetic calorimeter.
 Around 12500 sense wires stretched  between the DC endplates allow to obtain a track spatial resolution of $\sim$2~mm along the axis and better than
 $200  ~\mu \mbox{m}$ in the transverse plane. The accuracy on the decay vertex determination is $\sim$1~mm and the resolution of the  particle transverse momentum  is~$0.4~\%$~\cite{Ferrari2002}.
 The electromagnetic calorimeter (EMC) made by lead and scintillating fibers
is divided into a barrel and two end-caps,
has a readout granularity of $\sim (4.4 \times 4.4)$~cm$^2$,
for a total of 2440 cells arranged in five layers
covering $98\%$ of the solid angle.
It has energy and time resolution of
  $  \sigma(E) / E  = 5.7 \%/ \sqrt{E [\mbox{GeV}]}$,
  $  \sigma_{t} = 54 \mbox{ ps}/\sqrt{E [\mbox{GeV}] } \oplus 140 \mbox{ ps}$
for photons and electrons~\cite{Ambrosino2009}.

 The data acquisition  is enabled   by a two-level trigger system~\cite{Adinolfi:2002hs}.
   The first level trigger is a fast trigger with a minimal delay which starts 
   the acquisition at the front-end electronics. 
   It requires two local energy deposits above threshold (50 MeV on the barrel, 150 MeV on the end-caps).
   The trigger time is determined by the first particle reaching the calorimeter and is synchronized with the DA$\Phi$NE RF signal.
   
   The second level trigger 
   uses information from both the drift chamber and  the electromagnetic calorimeter.
 The trigger decision can be vetoed if the event is recognised as Bhabha scattering  or cosmic ray event.
 For control purposes these events are accepted and saved as  dedicated downscaled samples.

  The time interval between bunch crossings ($T_{bunch}= 2.715$~ns) is 
  smaller
  than the time spread 
  of the registered signals originating from $K_LK_S$ events that can reach 30-40~ns.
  The offline reconstruction procedure therefore has to determine the true bunch crossing time $T_0$ 
  for each event and correct all times related to that event accordingly.
  In the reconstruction algorithm the $T_0$ is determined by using 
  the EMC information. 
 In the studied channel, 
 since the $K_S$ decay  time is smaller than the $K_L$ interaction time in the calorimeter,
  the $T_0$ time has to be corrected in the offline analysis.

The data sample used for this analysis has been processed and filtered with the KLOE standard reconstruction software and the event classification procedure. 
The simulated data samples are based on the Monte Carlo (MC) GEANFI program~\cite{Ambrosino:2004qx}.



 \section{Measurement of $K_S\rightarrow \pi e \nu$ charge asymmetry}
  \label{sec::KS_selection}
 The charge asymmetry for the short-lived kaon is given~by:
 \begin{equation}
     A_S = \frac{N^+/\epsilon^+ - N^-/\epsilon^- }{N^+/\epsilon^+ + N^-/\epsilon^-},
	 \label{eq::as}
 \end{equation}
 where $N^+$ and $N^-$ are the numbers of observed $K_S \rightarrow \pi^- e^+ \nu$ and $K_S \rightarrow \pi^+ e^- \bar{\nu}$
 decays, respectively, 
 while $\epsilon^+$ and $\epsilon^-$ are the corresponding  efficiencies.
 Negative and positive charged pions
 interact differently in the detector material, 
 therefore  the
 efficiency is separately estimated for $\pi^- e^+ \nu$ and $\pi^+ e^- \bar{\nu}$
 final charge states. 

 \subsection{$K_S$ tagging}
The interaction of a $K_L$ meson in the calorimeter (crash) tags the presence of  a $K_S$ meson.
$K_L$ candidates must deposit  an energy $E_{clu}(crash) >$\cutClusterEnergy ~in the calorimeter in the polar angle range $40^\circ<\theta< 140^{\circ}$ and
not associated with a track from the DC.
Since the kaon velocity in the $\phi$ meson rest frame is well-defined ($\beta^*\sim0.22$), the requirement $\cutBeta$ is applied.
The $K_L$ direction obtained from the $K_L$ interaction coordinates in the calorimeter allows to determine
the $K_L$ momentum $\vec{p}_{K_L}$ with good precision, and hence the $K_S$ momentum:
$\vec{p}_{K_S} = \vec{p}_{\phi} - \vec{p}_{K_L}$. 

\subsection{Momenta smearing}
\label{sec::momenta_tuning}
In order to improve the MC simulation 
description of the experimental momentum resolution effects,
the reconstructed MC track
momentum components $p_i$ have been 
smeared using three Gaussian functions:
\begin{align}
	p_{i}^{new} & = p_i 
    \times 
    (1+\alpha_p) 
    \times 
    (1 +  \Delta \cdot \sum_{j=1}^{3} f_j \cdot G(0,\sigma_j) ), & (i=x,y,z)
\end{align}
where~$G(0,\sigma_j)$ is the Gaussian distribution with zero mean and standard deviation $\sigma_j$, $f_j$ is its amplitude,
while~$\Delta$ is the fractional 
uncertainty on the track curvature.

The momentum shift $\alpha_p$ and the Gaussian parameters 
are 
tuned on the $K_L \rightarrow \pi e \nu$ control sample (see Section~\ref{sec::KL_selection}). The fit yields $f_1= 96\%$,  $\sigma_1= 0.34$, $f_2=3.2\%$,  $\sigma_2= 9.74$, $f_3 =0.8\%$,
 $\sigma_3 = 71.2$ and $\alpha_p = 1.37\cdot 10^{-4}$.
 


\subsection{Event preselection} 
\label{sec:preselection}
\begin{figure}[h!]
	\centering
            \includegraphics[width=0.49\textwidth]{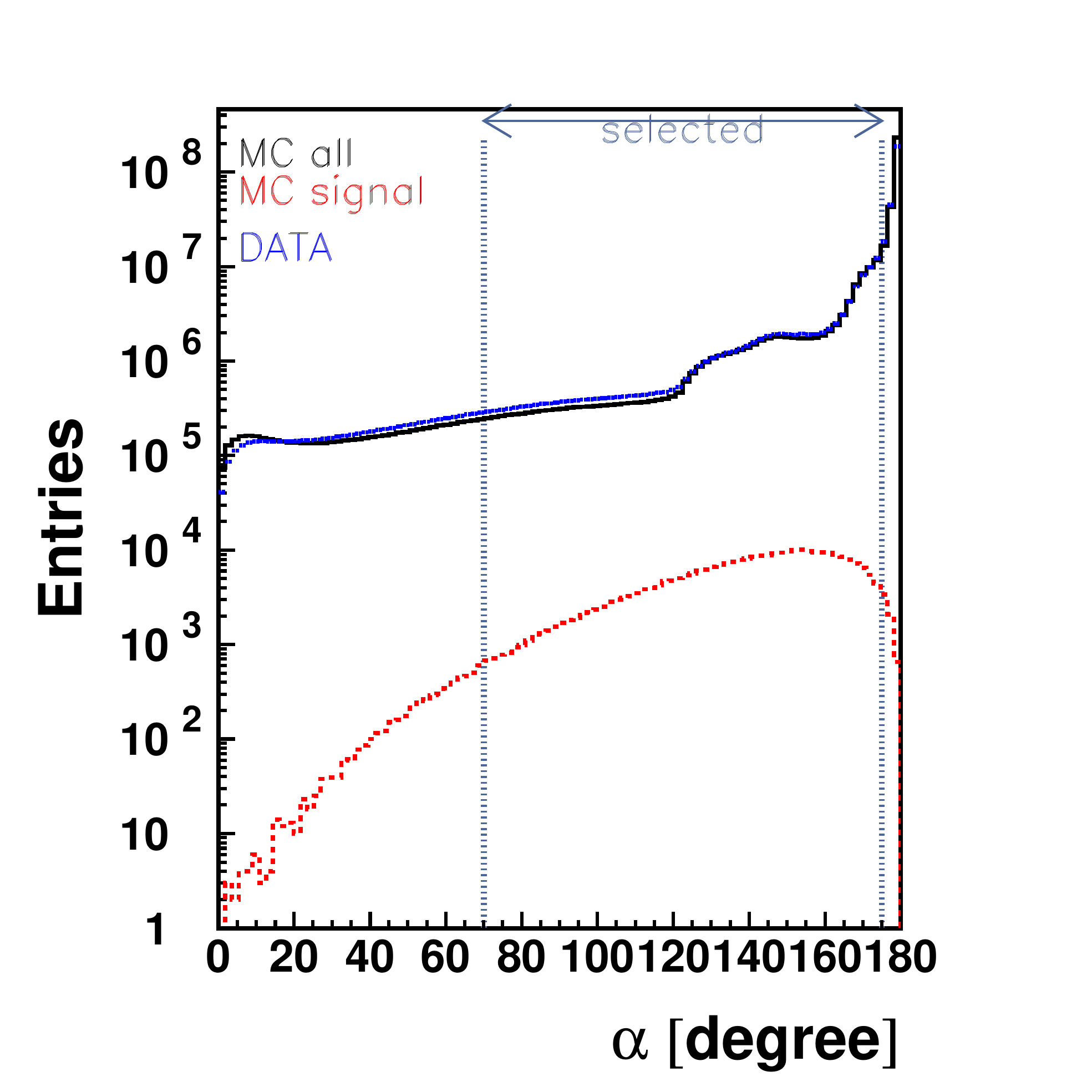}
            \includegraphics[width=0.49\textwidth]{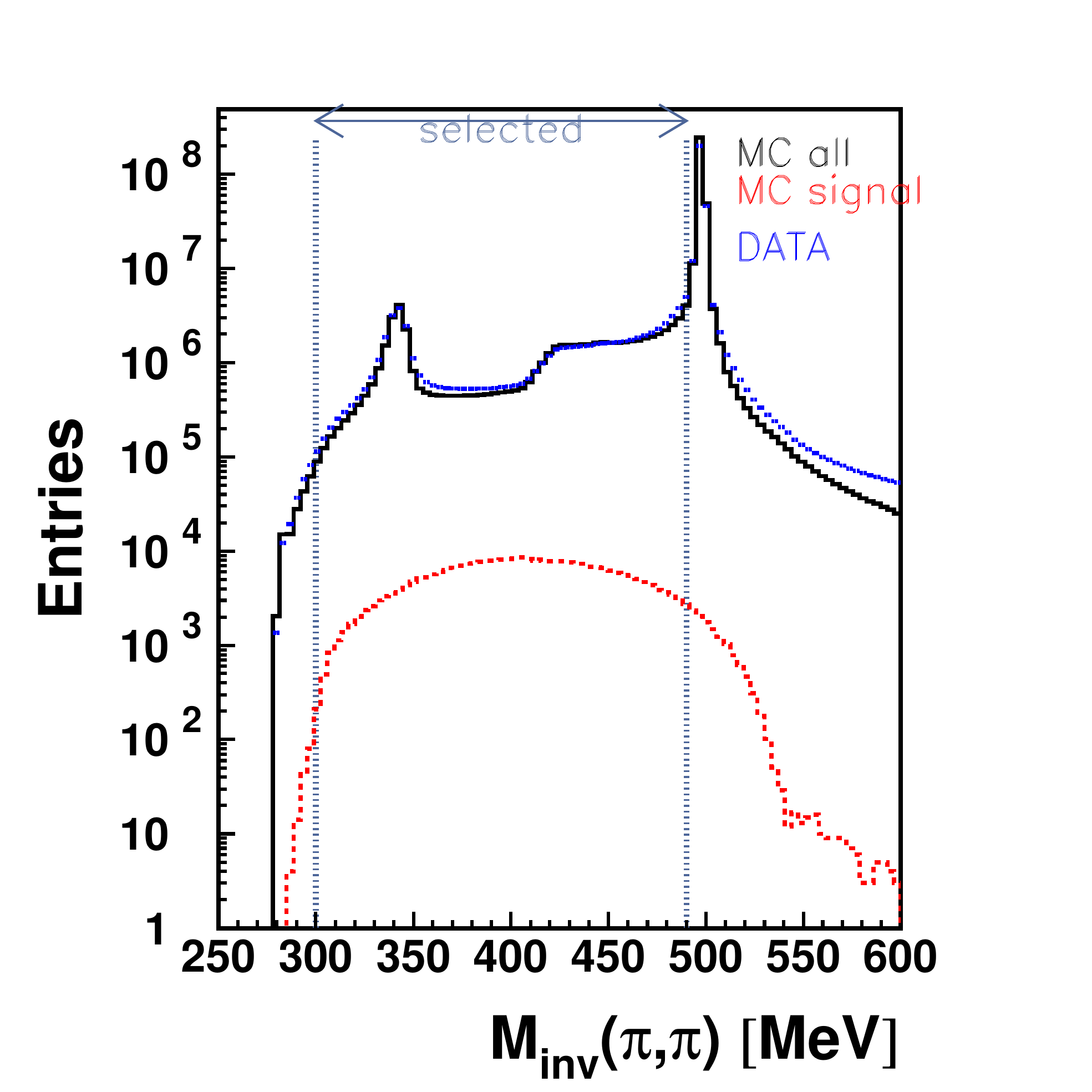}
    \caption{
	Left: Distribution of the $\alpha$ angle between charged secondaries in $K_S$ rest frame.
	Right:
Distribution of the invariant mass $M_{inv}(\pi,\pi)$ calculated under the assumption that both 
reconstructed tracks
are pions. In both figures black solid lines represent all simulated events, the red dashed 
lines
show simulated 
$K_S\rightarrow \pi e \nu$ signal events and blue points are data. Vertical dashed lines represent 
the
cuts described in the text.
    }
    \label{fig::rys_preselection}
\end{figure}

The selection of $K_S \rightarrow \pi e \nu$ decays starts with the reconstruction of a vertex 
 formed by two opposite curvature tracks 
 close to the $e^+e^-$ interaction point (IP) with $\cutRho$ and $\cutZ$,
 being
 $\rho_{vtx}$ and $z_{vtx}$ the transverse distance and the 
 longitudinal coordinate of the vertex, respectively.
 In the majority of the three-body decays of $K_S$ the angle between charged secondaries ($\alpha$) is contained in the (70$^{\circ}$, 175$^{\circ}$) range in the $K_S$ rest frame, 
 as shown in the left panel of Figure~\ref{fig::rys_preselection}.
Since the main source of background originates from the $K_S\rightarrow \pi^+ \pi^-$ decay, a cut on the invariant mass
under the assumption of both particles being charged pions is also applied 
($300~\hbox{MeV}<M_{inv}(\pi,\pi)<490~\hbox{MeV}$),
as indicated in  the right panel of Figure~\ref{fig::rys_preselection}.

Both tracks reconstructed in the drift chamber must be associated with  clusters in
the calorimeter by the Track to Cluster Association (TCA) procedure. 
This procedure extrapolates each track
from the last hit in the DC towards the calorimeter surface and determines
the impact point.

\subsection{Time of flight selection cuts}
\label{sec::TOFcuts}
\begin{figure}[htp]
            \includegraphics[width=0.32\textwidth]{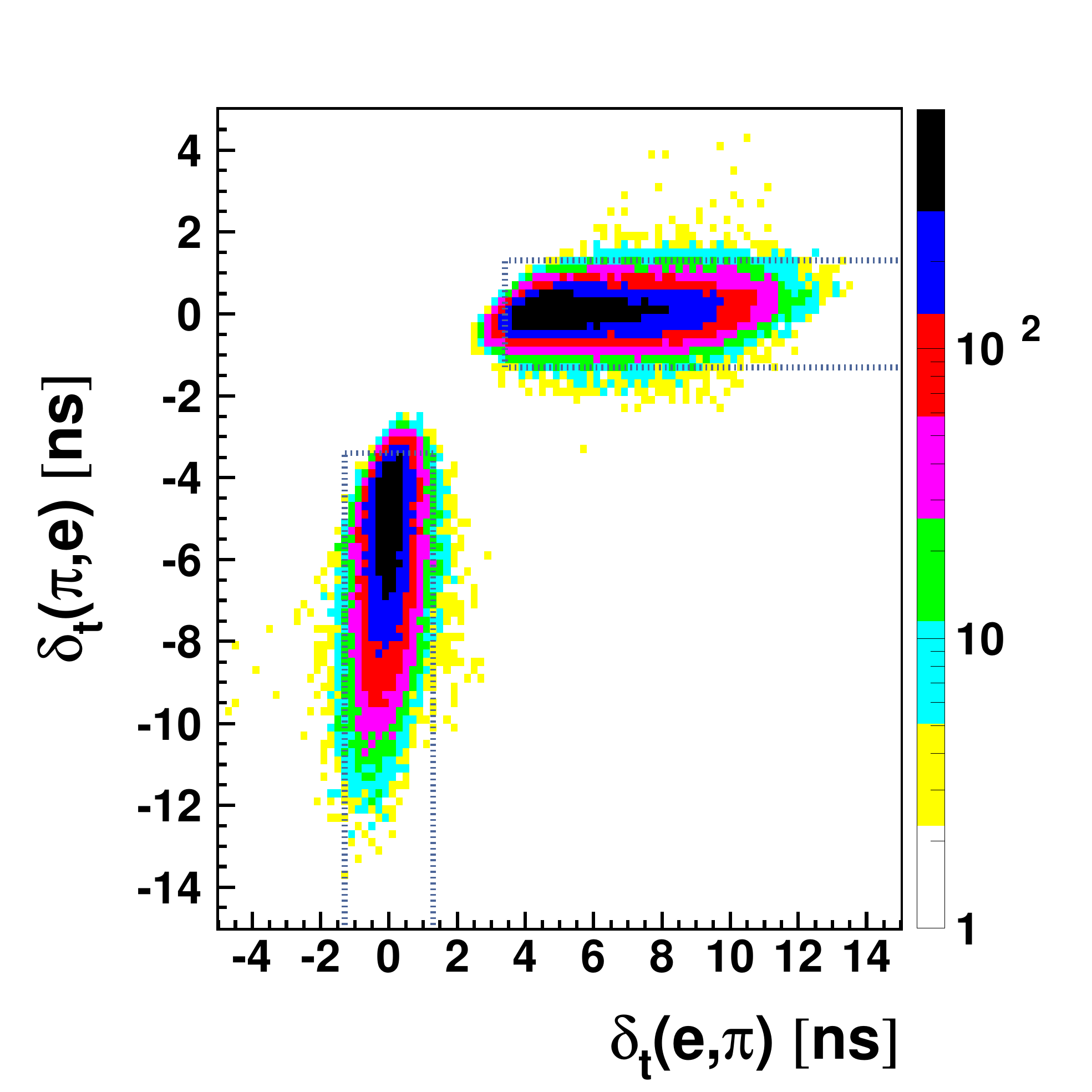}
            \includegraphics[width=0.32\textwidth]{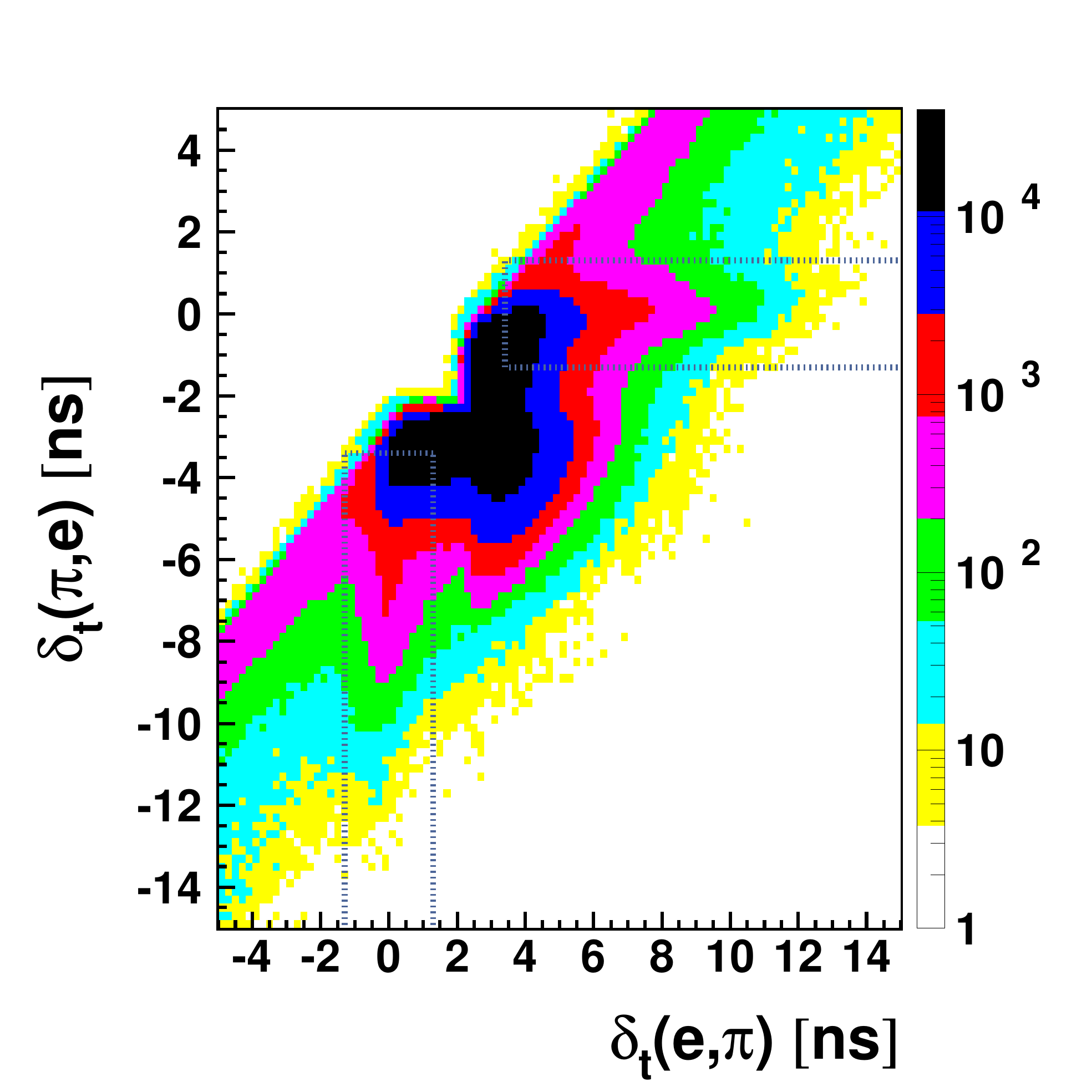}
            \includegraphics[width=0.32\textwidth]{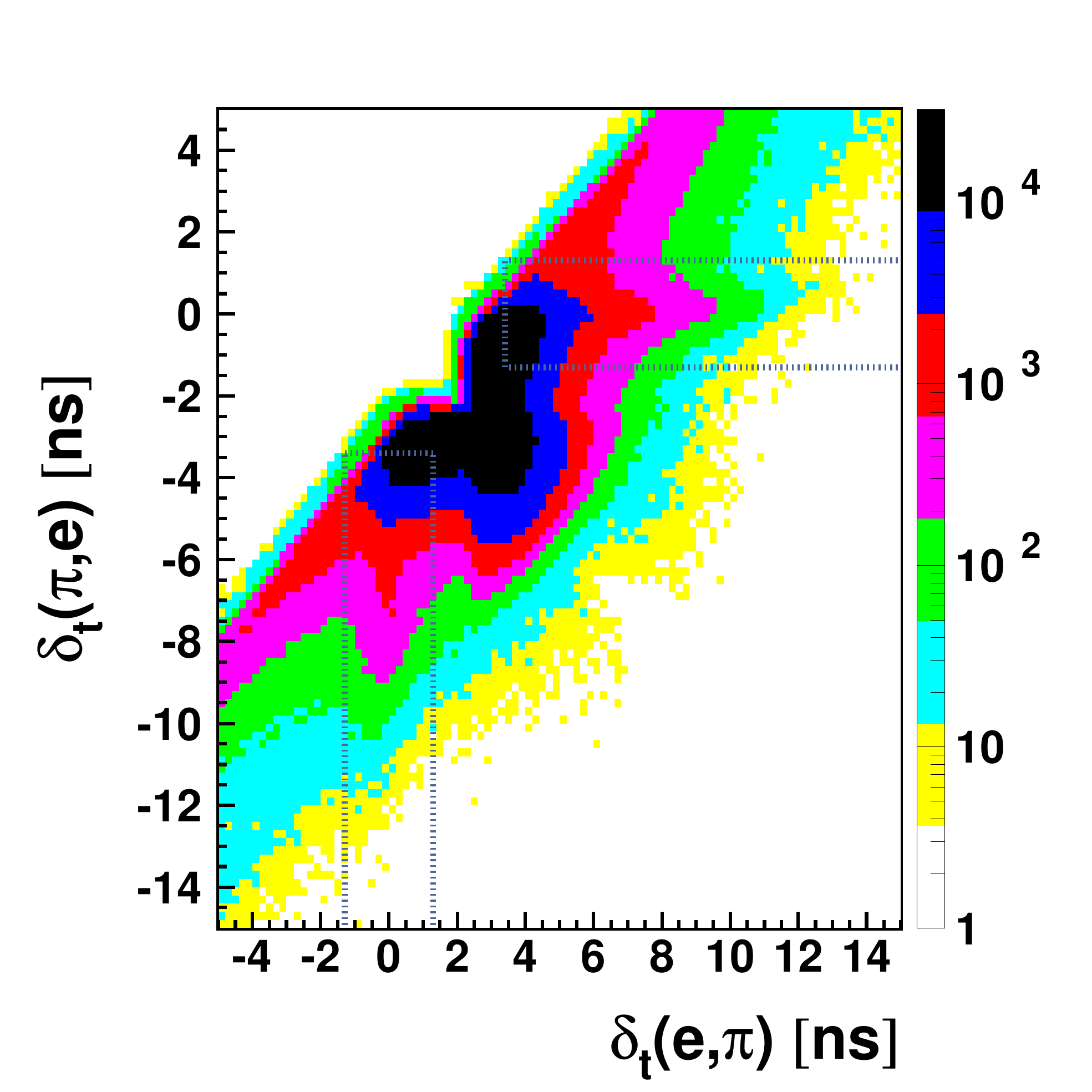}
    \caption{Distribution of TOF differences $\delta_{t} (\pi, e)$ vs $\delta_{t}(e ,\pi)$ 
        for simulated $K_{S}\rightarrow \pi e \nu$ events (left plot), all simulated events (center plot)
        and data (right plot). 
        The signal events are selected in the regions delimited by the dashed lines:
$\left(|\delta_{t} (e, \pi)| < 1.3 \ \mbox{ ns}, \delta_{t} (\pi, e) < -3.4  \mbox{ ns}\right)$ or 
$\left(\delta_{t}(e, \pi) > 3.4 \mbox{ ns},  |\delta_{t} (\pi, e) |< 1.3 \mbox{ ns}\right)$.
    }
    \label{fig::rys_tof2D}
\end{figure}

\begin{figure}[htp]
  \centering
            \includegraphics[width=0.49\textwidth]{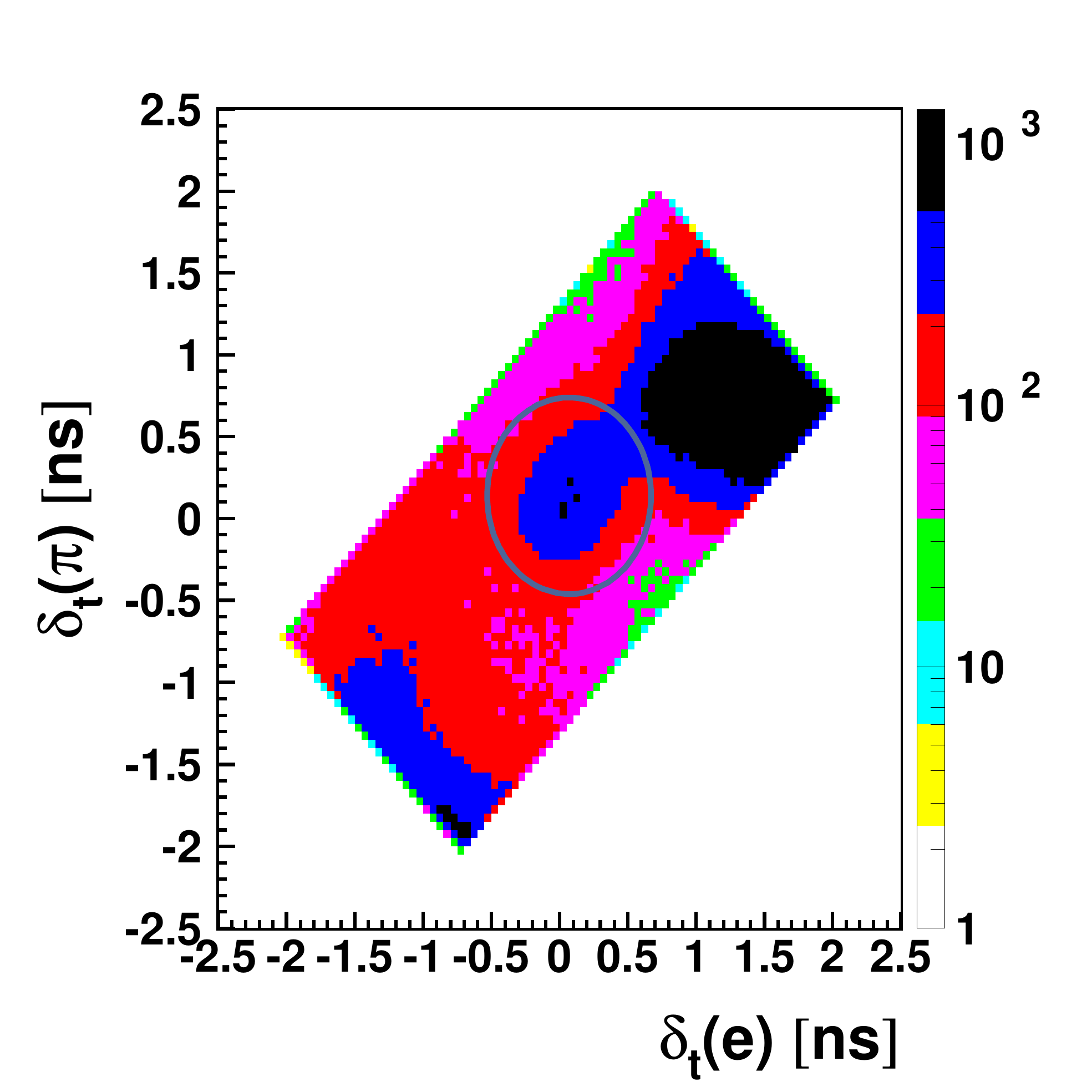}
            \includegraphics[width=0.49\textwidth]{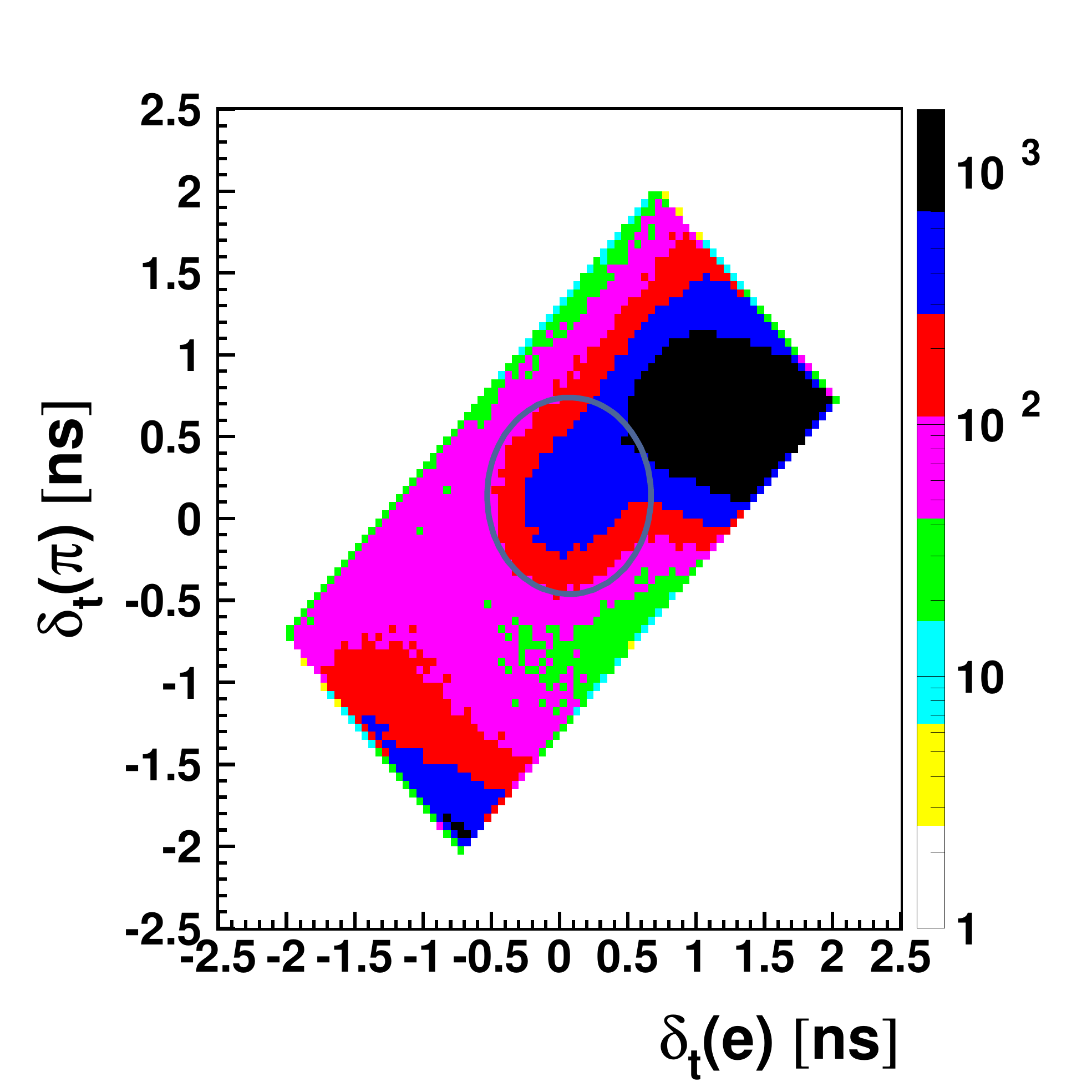}
      \\
            \includegraphics[width=0.49\textwidth]{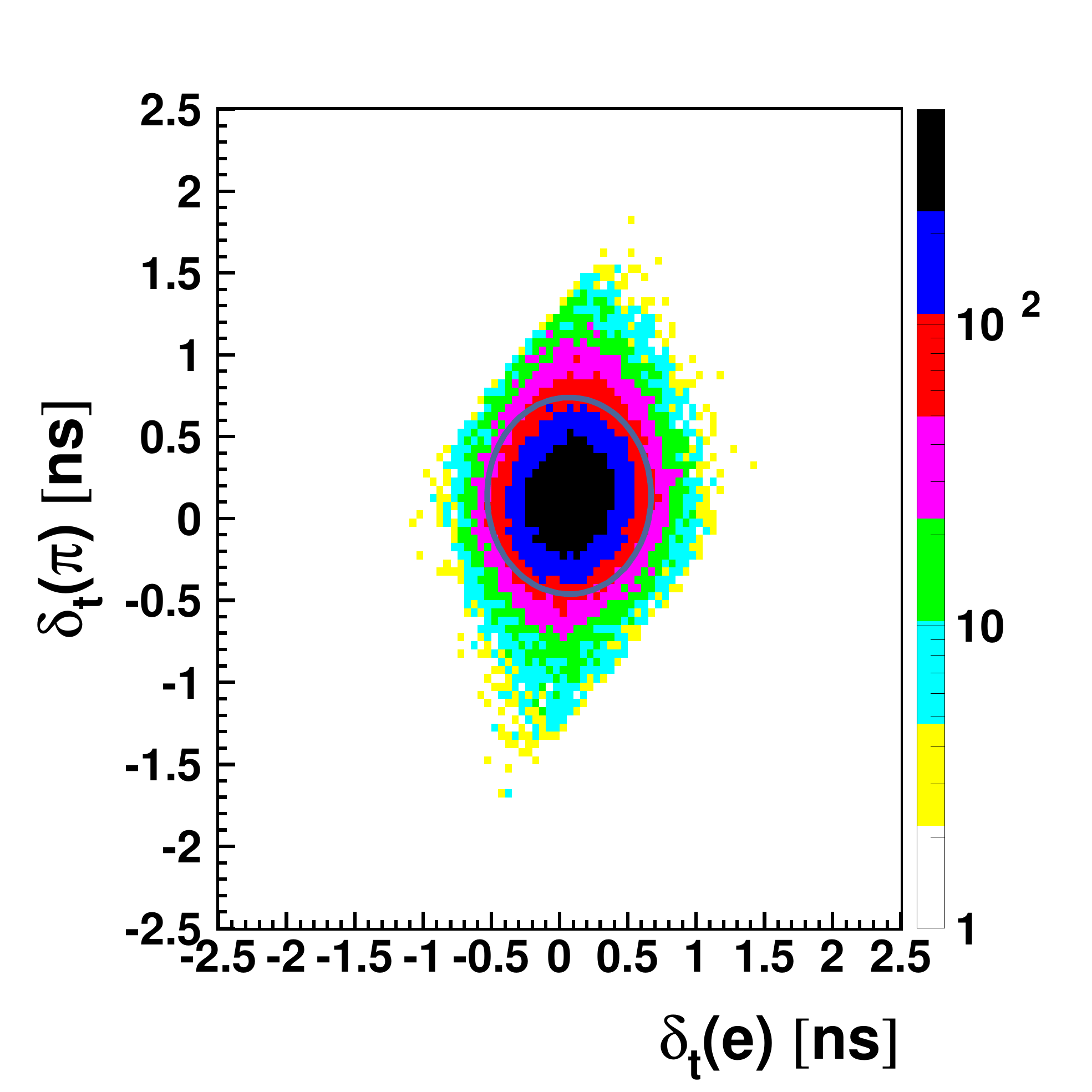}
            \includegraphics[width=0.49\textwidth]{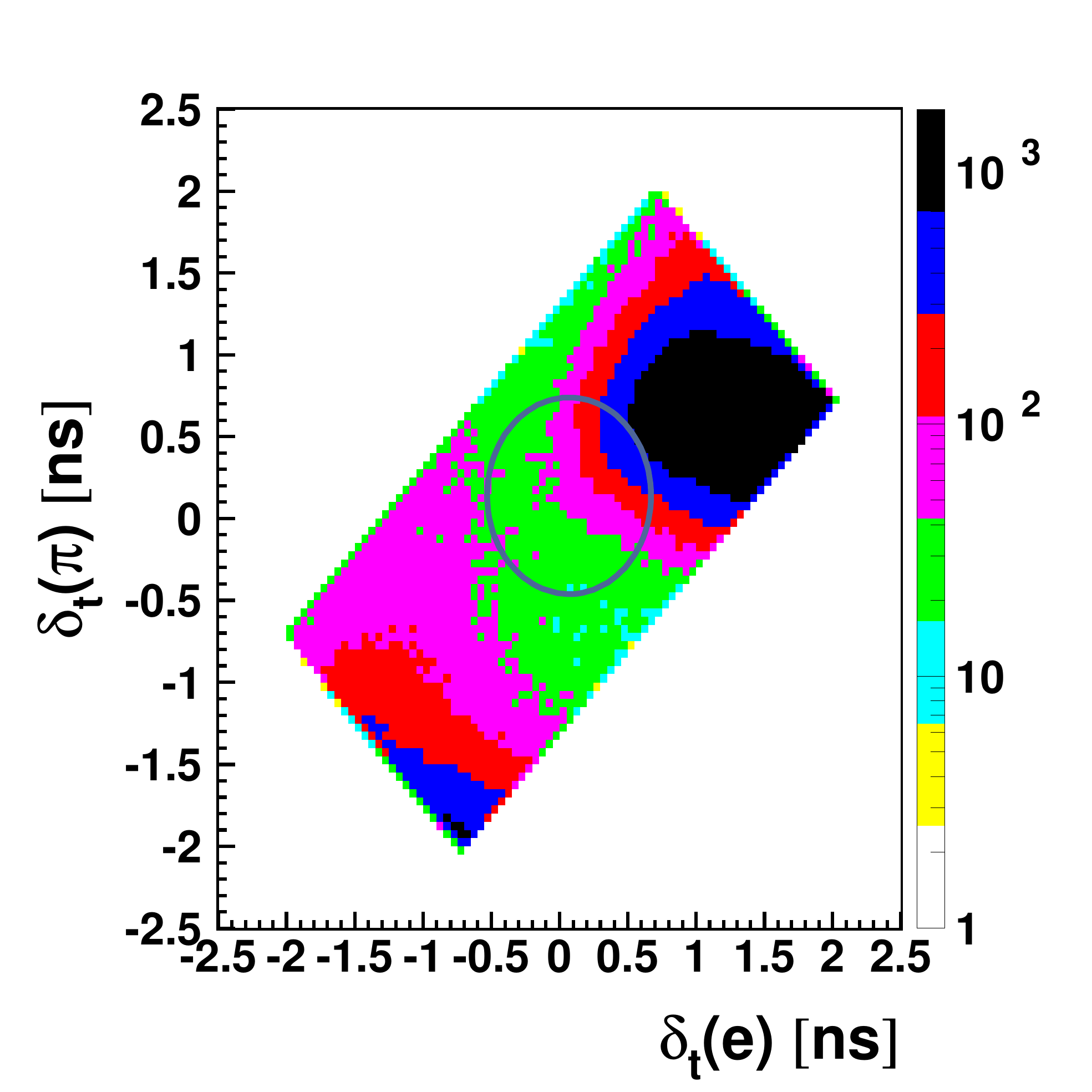}
      \caption{Distribution of the time differences
      $\delta_{t}(\pi)$
      vs
      $\delta_{t}(e)$
          for data events (top-left), all simulated events (top-right), simulated $K_{S} \rightarrow \pi e \nu$ events (bottom-left)
          and simulated background events (bottom-right). Events within the circle
          $\left[(\delta_t(e)-0.07~\mbox{ns})  \right]^2 + \left[ (\delta_t(\pi)-0.13~\mbox{ns}) \right]^2 
          = (0.6~\mbox{ns})^2$
          are retained for the analysis.
      }
      \label{fig::recalculated_dt}
\end{figure}

Further background reduction and final charged state ($\pi^{\pm} e^{\mp}$) identification is based on
 the difference $\delta_t(X)$ between the particle time of flight (TOF) from the $K_S$ decay vertex to the calorimeter ($t_{cl} - T_0$),
and the time calculated from the DC  measurement of  track length $L$ and particle momentum $p$
under the $m_X$  mass hypothesis\footnote{
The small $K_S$ decay time can be safely neglected here. In fact it identically cancels out in Equation \ref{eq::tof2d_epi}, 
while its average effect in the selection 
shown in Figure \ref{fig::recalculated_dt}
is accounted for by a small offset (of the order of $K_S$ lifetime) of the circle center with respect to the origin, with good agreement between data and MC.
}:
\begin{equation}
    \begin{aligned}
        \delta_{t}(X) &= (t_{cl} -T_0) - \frac{L}{c \cdot \beta(X)}, &  &
        \beta(X) =  \frac{p}{ \sqrt{p^2 + m_{X}^2 }}.
    \end{aligned}
    \label{tof_diff}
\end{equation}
Since at this stage the $\phi$ decay time ($T_0$) is 
not known with sufficient precision,
the following difference is introduced:
\begin{equation}
      \delta_{t}(X,Y) = \delta_{t}(X)_1 - \delta_{t}(Y)_2~,
      \label{eq::tof2d_epi}
  \end{equation}
where the mass hypothesis $m_{X(Y)}$ is used for track 1(2). 
Since for the correct mass 
assignments
the value of $\delta_{t}(X,Y)$ is close to zero, the condition $| \delta_{t}( \pi, \pi)| > 1.5 \mbox{ ns}$
 is applied
for further $K_S \rightarrow \pi^+ \pi^-$ 
rejection.
The remaining 
pairs of tracks 
are tested under pion-electron $\delta_{t}(\pi, e )$ and electron-pion $\delta_{t}(e,\pi)$ hypothesis (see Figure~\ref{fig::rys_tof2D}).
Once particle identification has been performed, the $T_0$ and the time differences $\delta_t(e)$ and $\delta_t(\pi)$
are  reevaluated accordingly.
Events are then selected within the circle in the $\delta_t(e) - \delta_t(\pi)$ 
plane as shown in Figure~\ref{fig::recalculated_dt}.

 The best separation between the signal and background components is obtained with the variable:
\begin{align}
    M^2(e)
	& =  \left[E_{K_S} - E(\pi)  - E_{\nu}\right]^2 -p^2(e), 
    \label{m2_def}
\end{align}
where
$E_{K_S}$ is computed from the kinematics of the two body decay $\phi\rightarrow K_S K_L$,
knowing the $\phi$-meson momentum (from Bhabha events) and the reconstructed $K_L$ direction,
$E(\pi)$ is evaluated from the measured track momentum in the pion hypothesis, and $E_{\nu} = |\vec{p}_{K_S} - \vec{p}(e) - \vec{p}(\pi) |$.
$M^2(e)$ is calculated according to the TOF particle identification. For the signal events  $M^2(e)$ peaks close to zero (see Figure~\ref{fig::normalization}).

\subsection{Signal extraction}
\label{sec::fit}
The signal yield is obtained by 
fitting the $M^2(e)$  distribution 
with a superposition of the corresponding
simulated distributions for signal and residual background components, with free normalizations,
separately for each final charge state,
and taking into account the statistical uncertainty of the Monte Carlo sample~\cite{Barlow:1993dm,Baker:1983tu}.
%
The remaining residual background components are:
\begin{itemize}
	\item the $K_S \rightarrow \pi^+\pi^-$ decays with one of the pion tracks not correctly reconstructed  and
           classified as an electron by the TOF algorithm (1.6\% of the sample after the fit, summing on the two final charge states);
   \item the $K_S \rightarrow \pi^+\pi^-$ decays where one of the pions decays into a muon before entering
            the drift chamber (18.7\%);
    \item radiative $K_S \rightarrow \pi^+\pi^-\gamma$ decays (2.5\%);
    \item other decays mainly originating from $\phi\rightarrow K^+ K^-$  (6.7\%)~.
\end{itemize}

\begin{figure}[h!]
    \centering
      \includegraphics[width=0.45\textwidth]{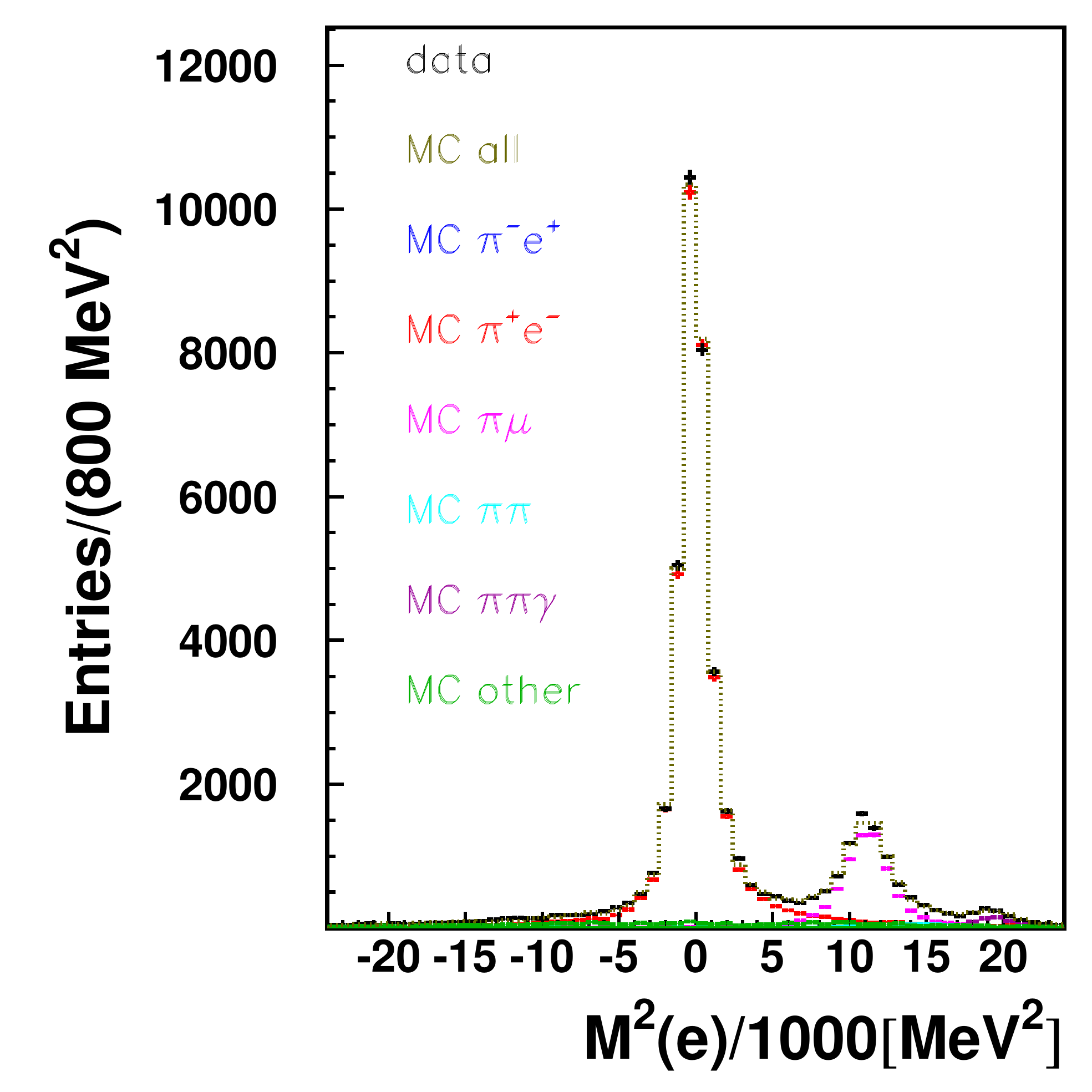}
      \includegraphics[width=0.45\textwidth]{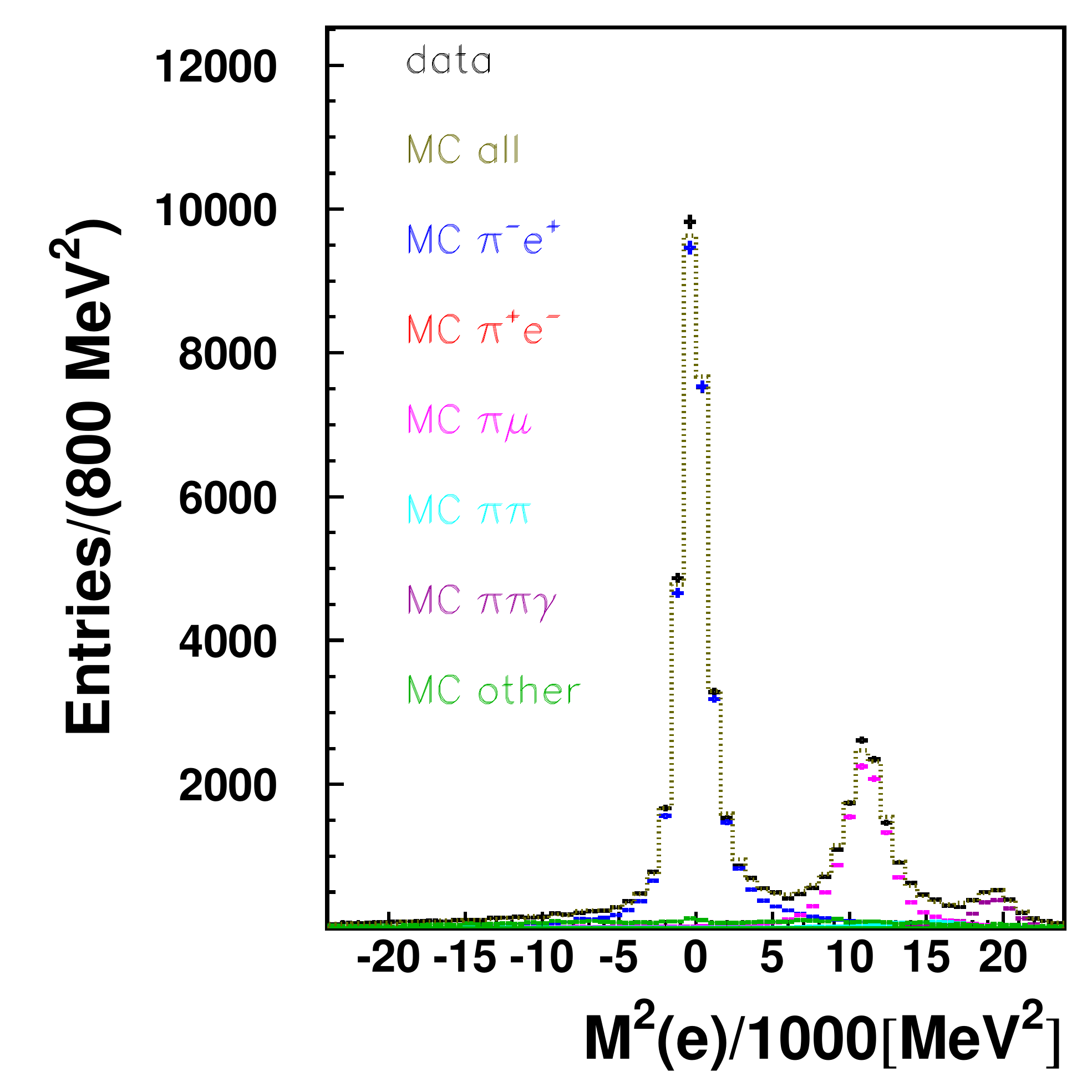}
      \\
      \includegraphics[width=0.45\textwidth]{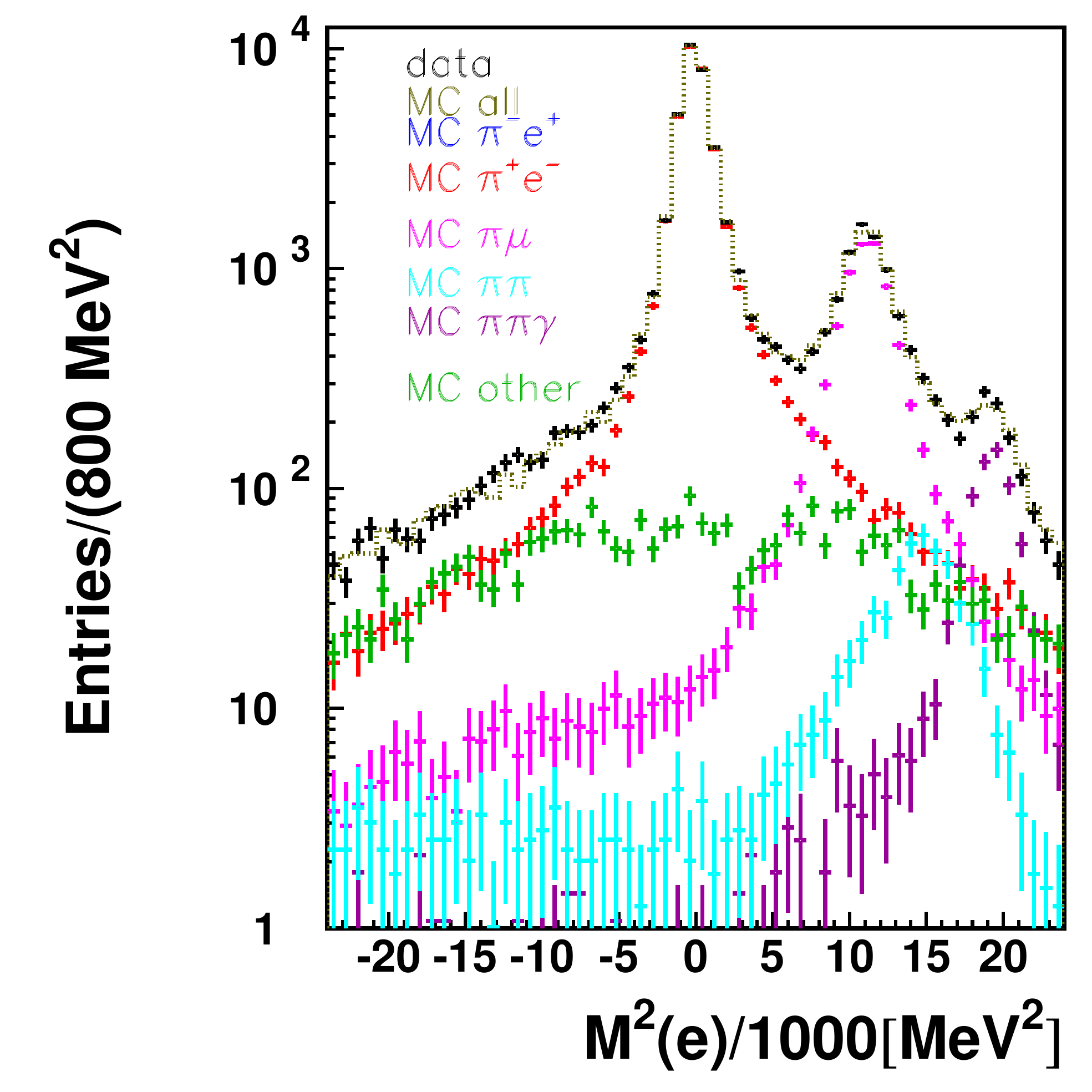}
      \includegraphics[width=0.45\textwidth]{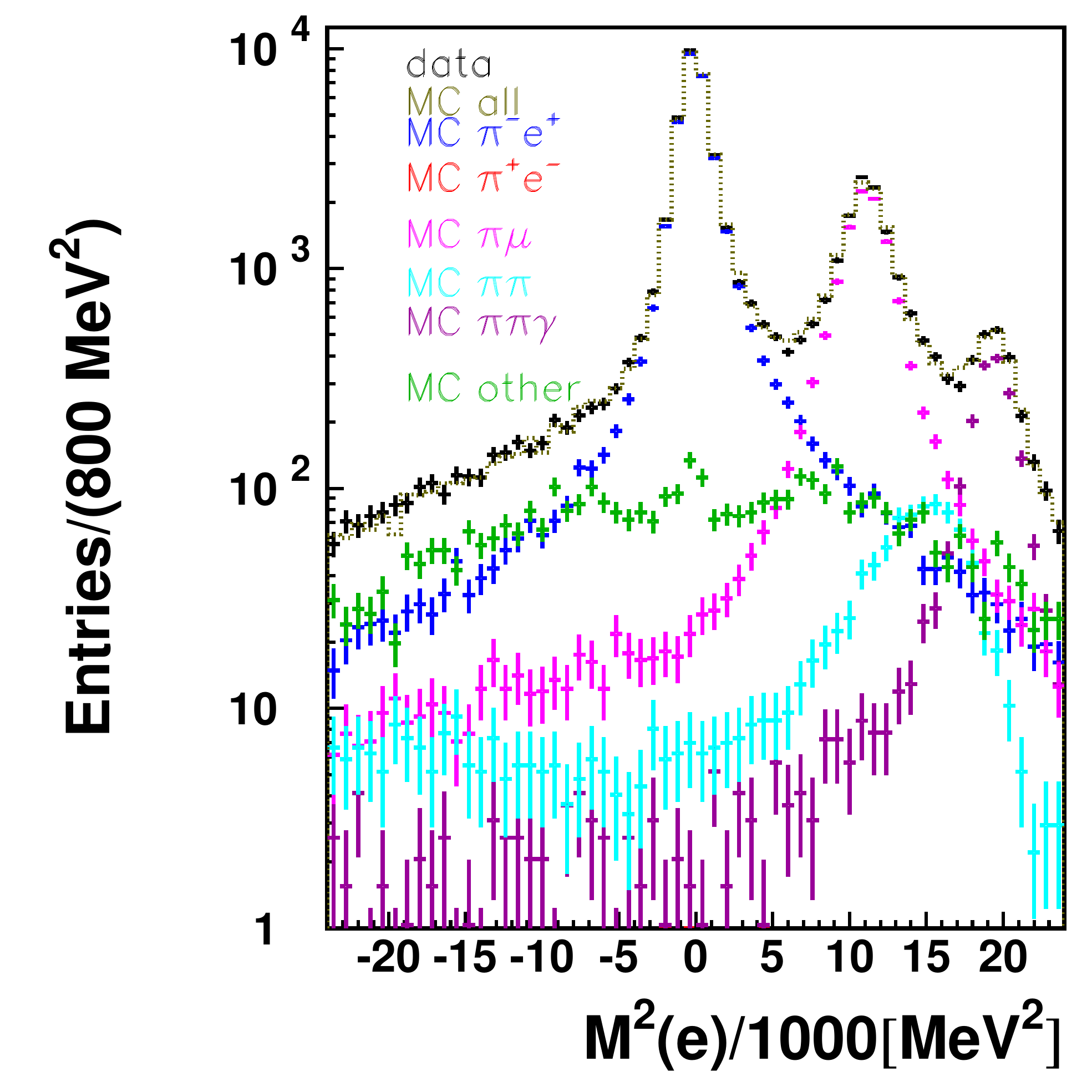}
      \\
      \includegraphics[width=0.45\textwidth]{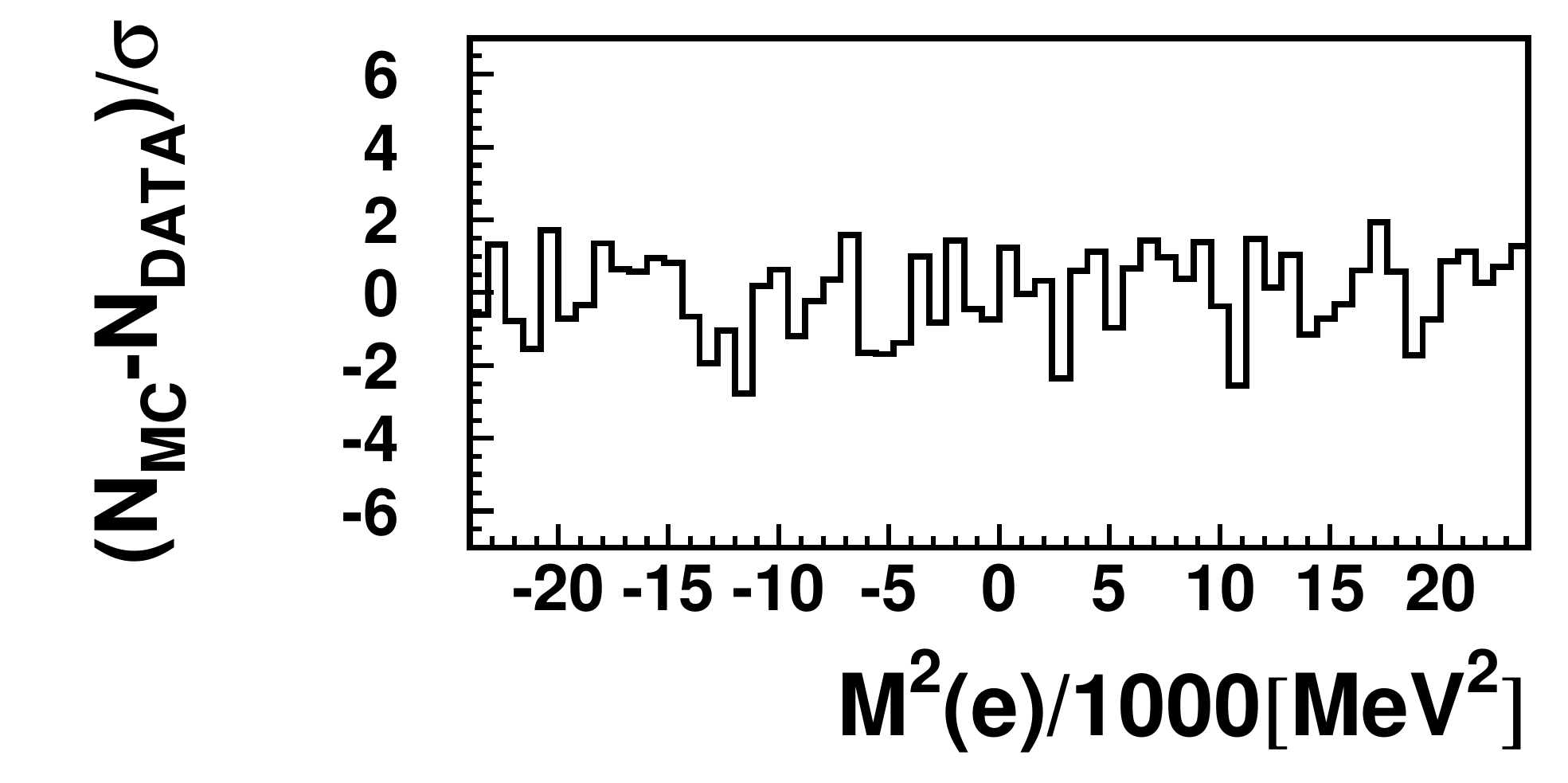}
      \includegraphics[width=0.45\textwidth]{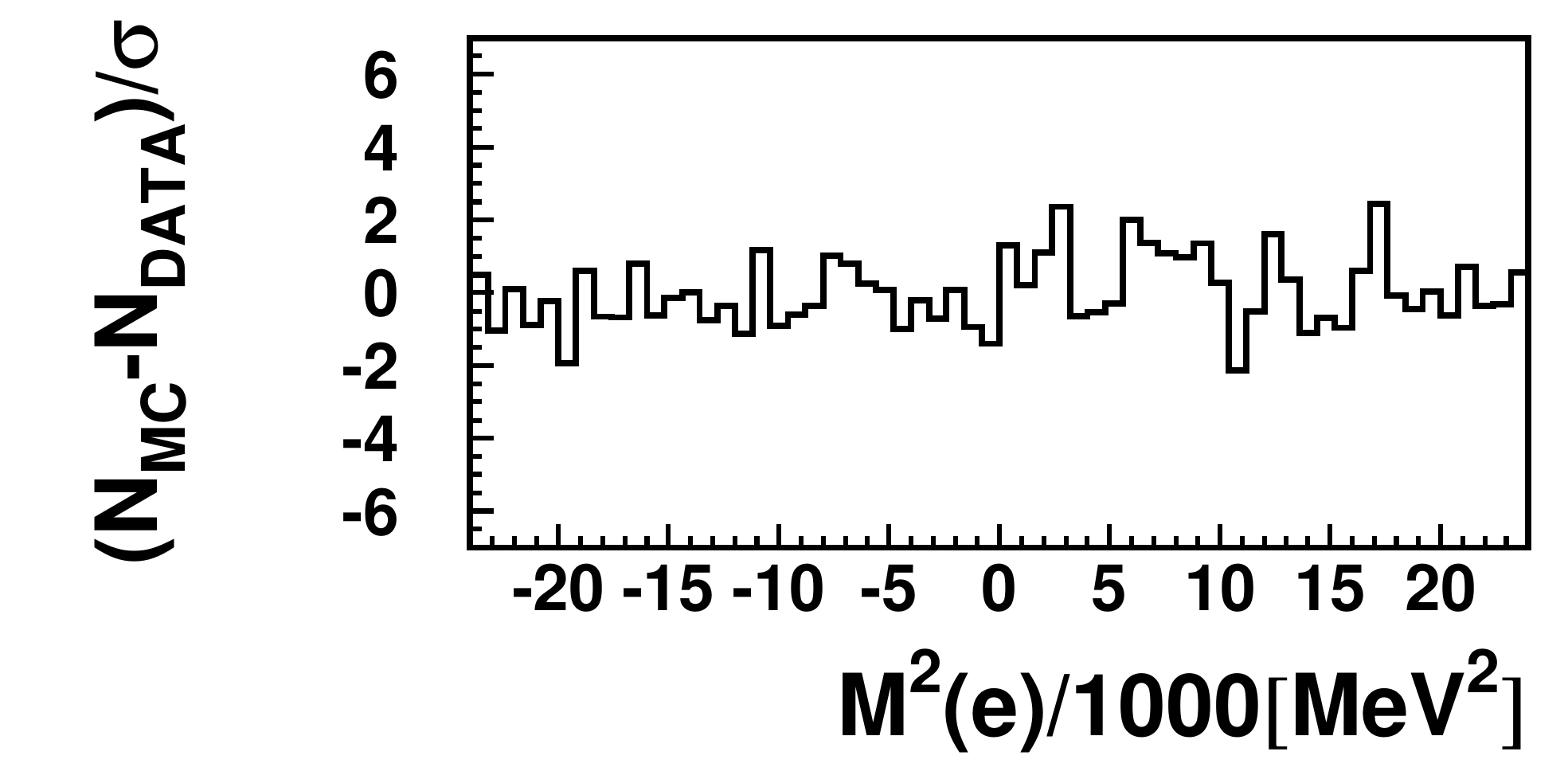}
      \caption{$M^2(e)$ distribution for
      data (black points) and MC simulation (dotted histogram)
        for both final charge states ($\pi^+ e^-$ 
        -- left side, $\pi^- e^+$ 
        -- right side) after the fit. The individual MC contributions are shown superimposed in the plots (colored points -- see legend in the plots).
      Bottom row: corresponding data-MC residual distributions after the fit. 
      }
      \label{fig::normalization}
\end{figure}

The result of the fit for the signal events is 
$\ksParEp \pm \ksParErrEp$  
for $K_S\to\pi^- e^+ \nu$ and 
$\ksParEn \pm \ksParErrEn$  for $K_S\to\pi^+ e^- \bar{\nu}$,  with total  $\chi^2/ndof=\ksChiAll$,  
summing on the two final charge states
(see Figure~\ref{fig::normalization}).


\section{$K_L \rightarrow \pi e \nu$ control sample selection}
\label{sec::KL_selection}
\label{sec::effi_control_sample}

A data sample of $K_L \rightarrow \pi e \nu$ decay, which is a dominant decay mode of $K_L$ meson,
is selected and used as a control sample. These events are tagged by the $K_S \rightarrow \pi^0 \pi^0$ decay\footnote{
Quantum interference effects in the double decay $K_L K_S \rightarrow \pi e \nu, \pi^0\pi^0$ are negligible in this specific case.}, 
identified by a total energy deposition in the calorimeter greater than 300~MeV, single photon deposit in the range from  20 to 300~MeV,  and the $\pi^0\pi^0$ invariant mass in the range from 390 to 600~MeV.
The estimated tag efficiency is $(60.0 \pm 0.3)\%$. 
No appreciable contamination is found from other $\phi$ meson decays and the  
beam-induced background is kept at the level of~1\%. 


Due to the different lifetimes  of  $K_S$ and $K_L$,
the vertex distribution of $K_L$ decays  is
weighted  to reproduce the $K_S$ distribution.
The weighting is performed  bin-by-bin  in the same $\rho_{vtx}$ -- $z_{vtx}$ 
acceptance region 
of 
the signal $K_S\rightarrow \pi e \nu$.
In this way the $K_L \rightarrow  \pi e \nu$ selected sample accurately mimicks the signal.


The events of the control sample are used to estimate directly from data the efficiencies for 
positive and negative pions. 
To this aim 
a single track selection scheme 
 is developed and applied, after
vertex reconstruction  and cuts on opening angle in $K_L$ rest frame and $M_{inv}(\pi,\pi)$,
as described in Section~\ref{sec:preselection}.

At this stage we require that at least one track 
reaches
the calorimeter with~TCA.  
For 
this 
track the 
$\delta_t (e)$ and $\delta_t (\pi)$ variables are constructed (see Equation~\ref{tof_diff}).
A pure sample of electrons (positrons) is then selected by requiring 
$ \left[ (\delta_t(e) - 0.07~\mbox{ns})/1.2~\mbox{ns} \right]^2 + \left[ (\delta_t(\pi) +4~\mbox{ns} )/3.2~\mbox{ns} \right]^2$ $ <1$. 
%
Assuming the other track is a $\pi^+$ 
(or a $\pi^-$),
we can test if it is associated to a calorimeter cluster to obtain the TCA efficiency $\epsilon_{TCA}^{KL \ DATA}(\pi^{\pm})$, separately for negative and positive pions.
For $e^{\pm}$ tracks we use the MC simulation to estimate the corresponding efficiencies, $\epsilon_{TCA}^{KS \ MC}(e^{\pm})$.
When the pion is associated to a cluster, 
then we can test if both tracks satisfy 
the TOF selection cuts described in Section
\ref{sec::TOFcuts}
in order to obtain directly from the control sample (in this case without using MC) the combined efficiencies 
$\epsilon_{TOF}^{KL \ DATA}(\pi^{\pm} e^{\mp})$.
\\
The different tagging conditions for $K_S$ and $K_L$ samples are taken into account 
by correcting $\epsilon_{TCA}^{KL \ DATA}(\pi^{\pm})$ and $\epsilon_{TOF}^{KL \ DATA}(\pi^{\pm} e^{\pm})$ 
for the ratio of the same efficiencies obtained from MC for $K_S$ and $K_L$ samples,
$\epsilon_{TCA}^{KS \ MC}(\pi^{\pm}) / \epsilon_{TCA}^{KL \ MC}(\pi^{\pm})$ and 
$\epsilon_{TOF}^{KS \ MC}(\pi^{\pm} e^{\mp}) / \epsilon_{TOF}^{KL \ MC}(\pi^{\pm} e^{\mp})$, respectively.

\section{Efficiency determination}
\label{sec::effi}
The total $K_S\rightarrow \pi e \nu$  selection efficiency is estimated as~follows:
\begin{align}
	\epsilon & =  \epsilon_{TEC}
	 \cdot
            \epsilon_{TAG} \cdot
            \epsilon_{ANA},
    \label{eq:effi_tot}
\end{align}
where $\epsilon_{TEC}$  stands for
trigger and event classification efficiency, while $ \epsilon_{TAG}$ and $\epsilon_{ANA}$ 
denote tagging and analysis efficiencies, respectively.

The analysis efficiency $\epsilon_{ANA}$  can be expressed in~turn as a product
of four contributions:
\begin{itemize}
	\item kinematical cuts ($\epsilon_{KC}$):  cuts 
		on reconstructed vertex fiducial volume, opening angle $\alpha$,
		and $M_{inv}(\pi,\pi)$ (see Section~\ref{sec:preselection});
	\item Track to Cluster Association algorithm ($\epsilon_{TCA}$);
	\item Time of Flight cuts ($\epsilon_{TOF}$); 
	\item  fit range ($\epsilon_{FR}$) of the  $M^2(e)$ variable.
\end{itemize}

The efficiency $\epsilon_{TEC}$ is evaluated using  downscaled
minimum-bias data samples without event classification and background rejection filters. 
The estimation of  $\epsilon_{TAG}$, $\epsilon_{KC}$ and $\epsilon_{FR}$ are based on MC simulation;
$\epsilon_{TCA}$ and $\epsilon_{TOF}$  
are determined using the $K_L \rightarrow \pi e \nu$ control sample with the 
method described in Section~\ref{sec::effi_control_sample};
$\epsilon_{TCA}$ consists of the product 
of $\epsilon_{TCA}(\pi^{\pm})$ and $\epsilon_{TCA}(e^{\mp})$,
the first evaluated from the control sample and the second from MC:
 \begin{align}
	 \epsilon_{TCA} & =  \epsilon_{TCA}^{KS}(\pi) \times \epsilon_{TCA}^{KS}(e) =  \epsilon_{TCA}^{KL \ DATA}(\pi) \times \frac{\epsilon_{TCA}^{KS \ MC}(\pi)}{\epsilon_{TCA}^{KL \ MC}(\pi)} \times  \epsilon_{TCA}^{KS \ MC}(e),
 \end{align}
 while $\epsilon_{TOF}$ 
is determined using the $K_L\rightarrow\pi e\nu$ data control sample with events in which both tracks are associated to a calorimeter cluster and identified:
\begin{align}
    \epsilon_{TOF} & 
    = \epsilon_{TOF}^{KL \ DATA}(\pi e) \times \frac{\epsilon_{TOF}^{KS \ MC}(\pi e)}{\epsilon_{TOF}^{KL \ MC}(\pi e)}. 
\end{align}
Both $\epsilon_{TCA}$ and $\epsilon_{TOF}$ have been corrected for the different tagging conditions of the control sample.
\par
The total efficiency is 
$(\ksEffiEp  \pm \ksEffiErrEp) \%$ and 
$(\ksEffiEn \pm \ksEffiErrEn) \% $, for
$K_S \rightarrow \pi^- e^+ \nu$ and $K_S \rightarrow \pi^+ e^- \bar{\nu}$, respectively.
The evaluated efficiencies for the different analysis steps are presented in Table~\ref{tab_twopi0_ks_effi}.
\begin{table}[h!]
    \caption{Efficiencies (\%) for the different analysis steps.}
    \label{tab_twopi0_ks_effi}
    \begin{center} 
        \begin{tabular}{|C{0.45\textwidth}|C{0.2\textwidth}|C{0.2\textwidth}|} \hline
		Efficiency $(\%)$			& $K_S \rightarrow \pi^- e^+ \nu$			& $K_S \rightarrow \pi^+ e^- \bar{\nu}$         \\ \hline \hline
        trigger and event classification ($\epsilon_{TEC})$	& $\ksEffiEpFilfo \pm \ksEffiErrEpFilfo$		& $\ksEffiEnFilfo \pm \ksEffiErrEnFilfo$ \\ \hline 
        $K_S$ tagging	$(\epsilon_{TAG})$			& $\ksEffiEpTag  \pm \ksEffiErrEpTag$		& $\ksEffiEnTag   \pm \ksEffiErrEnTag $  \\ \hline
        kinematical cuts $(\epsilon_{KC})$			& $\ksEffiEpDC  \pm  \ksEffiErrEpDC$		& $\ksEffiEnDC    \pm \ksEffiErrEnDC $   \\ \hline
        Track to Cluster Association	$(\epsilon_{TCA})$	& $\ksEffiEpTCA  \pm \ksEffiErrEpTCA$		& $\ksEffiEnTCA   \pm \ksEffiErrEnTCA$   \\ \hline
        Time of Flight	$(\epsilon_{TOF})$			& $\ksEffiEpTOF  \pm \ksEffiErrEpTOF$		& $\ksEffiEnTOF   \pm \ksEffiErrEnTOF$   \\ \hline
        Fit range  $(\epsilon_{FR})$  & $\ksEffiEpOther  \pm \ksEffiErrEpOther$	& $\ksEffiEnOther \pm \ksEffiErrEnOther$ \\
        \hline
    \end{tabular}
    \end{center}
\end{table}

 Using these efficiencies in eq.~\ref{eq::as}
 the result for $A_S$ is: 
 \begin{equation}
	 A_S = (\ksAs \pm \ksAsErrStat_{stat}) \times 10^{-3}.
     \label{eq::as_stat}
 \end{equation}

\section{Systematic uncertainty }
\label{sec::systematic}

In order to estimate the contributions to the systematic uncertainty,  the full analysis chain 
is repeated  varying all the analysis cut values of selection variables by $+$/$-$ an amount 
comparable with their experimental resolution.
These variations probe the level of accuracy of the MC simulation; a data-MC disagreement could be due both to an imperfect detector simulation and/or 
to a bias in the estimate of the background induced by the machine or 
from other physical processes.
The contributions from the stability of  $M^2(e)$ distribution fit,
   momenta smearing, trigger and event classification procedures  are also estimated. 
   Unless differently specified in the following, each contribution is calculated as the absolute deviation from the nominal result (\ref{eq::as_stat}) averaged on the two 
  $+$/$-$ variations.
The stability of the $A_S$ result 
is 
also 
checked  along the running period and
against larger variations of the cut values.
The resulting values for $A_S$ do not exhibit any 
anomaly; their  behaviour is monotone or smooth.
 \par
The systematic uncertainties are classified into the following groups  (see Table~\ref{tab::systScan::all}):
 \begin{itemize}
     \item  Trigger and event classification:
         \begin{itemize}
             \item 
         Systematic effects originating from  the trigger and the event classification procedure 
         are estimated in prescaled data samples.
         The analysis of the prescaled samples follows the standard analysis
			 chain. The systematic contribution ($\sigma_{TEC}$) is estimated to be $0.28\times 10^{-3}$. 
         \end{itemize}
     \item  Tagging and preselection: 
         \begin{itemize}
             \item 
                 The $K_L$ deposited energy cut
                 is changed to the values $E_{clu}(crash)$ $= \{95,$ $105, 110,$ $115, 150,$ $200\}$~MeV. 
                 The stability of the result is checked within this range.
                 The systematic uncertainty is evaluated by changing the cut by $\pm 5$~MeV.
             \item
                 The $\beta^*$ interval is 
                 enlarged or shrunk 
                 by $0.02$ ($1 \sigma$) on each side ($0.18\mp0.02 < \beta^* < 0.27\pm0.02$).
                 The stability of the result is checked up to a variation of $\pm 5 \sigma$.
             \item 
                 The $z_{vtx}$ and $\rho_{vtx}$ cuts for the reconstructed $K_S \rightarrow \pi e \nu$ decay vertex position 
                 are each independently varied by $\pm 0.2$~cm ($\pm 1 \sigma$).
                 The stability of the result is checked against a variation of $\pm 5 \sigma$. 
				\item
                The range of the opening angle $\alpha$ of the charged secondaries in the $K_S$ rest frame is enlarged or shrunk by $2^{\circ}$ ($1 \sigma$) on each side ($70\mp 2^{\circ} < \alpha < 175\pm2^{\circ}$).
                The stability of the result is checked up to a variation of $\pm 5 \sigma$ with the constraint of the upper bound not exceeding $180^{\circ}$.
                \item
                The $M_{inv}(\pi,\pi)$ interval is enlarged or shrunk by 1 MeV ($1 \sigma$) on each side ($300\mp 1 ~\hbox{MeV} < M_{inv}(\pi,\pi) < 490\pm 1~\hbox{MeV}$).
                The stability of the result is checked up to a variation of $\pm 5 \sigma$.
         \end{itemize}
     \item  Time of flight selection:
         \begin{itemize}
         \item
         The $|\delta_t (\pi,\pi)|$ cut is varied 
         by $\pm 0.1$~ns. The stability of the result is checked up to a variation of $\pm 0.4$~ns.
         \item 
         The regions for the selection of the signal in the $\{ \delta_t (e,\pi),\delta_t (\pi,e)\}$ plane are enlarged or shrunk by varying the cuts of $\pm 0.1$~ns~
         ( 
         $
         [
         |\delta_{t} (e, \pi)| < 1.3\pm0.1 \ \mbox{ ns}$~,$\delta_{t} (\pi, e) < -3.4 \pm0.1 \   \mbox{ ns}
         ]
         $ 
         or 
$
[
\delta_{t}(e, \pi) > 3.4 \mp 0.1\mbox{ ns}$~,  $|\delta_{t} (\pi, e) |< 1.3 \pm 0.1 \mbox{ ns}
]
$ ). The stability of the result is checked up to
variations of $\pm 0.4$~ns.
         \item
         The circular region for selection of the signal in the $\{\delta_t (e),\delta_t (\pi)\}$ plane is enlarged or shrunk by varying its radius of $\pm 0.1$~ns.
             The stability of the result is checked for variations ranging from $-0.3$~ns to $+0.4$~ns.
         \end{itemize}
     \item Momenta smearing: 
         \begin{itemize}
             \item 
             The $K_L\rightarrow \pi e \nu$ control sample is divided into ten, equal in luminosity subsamples.      
The momenta smearing parameters are tuned separately for each subsample.
From the standard deviation of the results the systematic contribution ($\sigma_{MS}$)
          is estimated to be $0.58 \times 10^{-3}$. 
     \end{itemize}
     \item  Fit procedure:
         \begin{itemize}
             \item 
                 The systematic uncertainty from the histogram bin width $\sigma_{HBW}$ 
is determined by varying the bin width 
from $0.8$    to $1.6$~MeV$^2/1000$ (this variation corresponds 
to the  $M^2(e)$ resolution evaluated from MC). $\sigma_{HBW}$ is estimated to be $0.61 \times 10^{-3}$. The stability of the result is checked for variations of the bin width from $2\sigma$ to $5\sigma$.
\item
The systematic uncertainty from the fit range is evaluated by varying 
it from $[-24:24]$~MeV$^2/1000$ 
to $[-28:28]$~MeV$^2/1000$ or  $[-20:20]$~MeV$^2/1000$.
The stability of the fit procedure is checked for histogram ranges 
from $[-36:36]$~MeV$^2/1000$ to $[-12:12])$~MeV$^2/1000$, 
while keeping the 
nominal bin size.
%
%
         \end{itemize}
 \end{itemize}
 The total systematic uncertainty is estimated as the sum in quadrature of the contributions listed above 
 and reported in Table~\ref{tab::systScan::all}.
 \par As a cross-check, the $A_L$ value for the  $K_L \rightarrow \pi e \nu$  control  sample 
is  determined following the same analysis steps as for $A_S$.
The result  $A_L = (1.7 \pm 2.7_{stat}) \times 10^{-3}$ is
consistent with the KTeV measurement~\cite{AlaviHarati:2002bb}.

 \begin{table}[h!]
 \caption{ 
     Summary of contributions to the systematic uncertainty on $A_S$.
}
  \label{tab::systScan::all}
 \centering 
 \begin{tabular}{|C{0.4\textwidth}|C{0.25\textwidth}|C{0.25\textwidth}|} \hline
 \multicolumn{2}{|c|}{Contribution}
     &  Systematic uncertainty ($10^{-3}$) \\ \hline \hline 
      Trigger and event classification & $\sigma_{TEC}$     & 0.28\\ \hline
Tagging and preselection & $E_{clu}(crash)$ & 0.55 \\ \hline 
     " & $\beta^{*}$  & 0.67 \\ \hline 
     " & $z_{vtx}$    & 0.01\\ \hline 
     " & $\rho_{vtx}$ & 0.05 \\ \hline 
     " & $\alpha$   & 0.46 \\ \hline 
     " & $M_{inv}(\pi,\pi)$ & 0.20 \\ \hline 
     Time of flight selection & $\delta_t (\pi,\pi)$ &  0.71 \\ \hline 
     " & $\delta_t (e,\pi)$ vs  $\delta_t (\pi,e)$   & 0.87 \\ \hline 
     " & $ \delta_t (e)$ vs  $\delta_t (\pi)$  & 1.82 \\ \hline 
     Momenta smearing  & $\sigma_{MS}$         &  0.58\\  \hline 
     Fit procedure & $\sigma_{HBW}$ &  0.61  \\  \hline
     " & Fit range & {\centering  0.49}\\ \hline                           
 \hline
 \multicolumn{2}{|c|}{Total}
     &  $\ksAsErrSyst$ \\ \hline
 \end{tabular}
 \end{table}

\section{Results}
\label{sec::summary}
The result for the $K_S \rightarrow \pi e \nu$ charge asymmetry is:
\begin{equation}
    \asresult,
\end{equation}
 consistent with the previous determination on an independent data sample~\cite{Ambrosino:2006si}
and improving the  statistical accuracy by almost a factor of two.  

\begin{figure}[h!]
    \centering
      \includegraphics[width=1.0\textwidth]{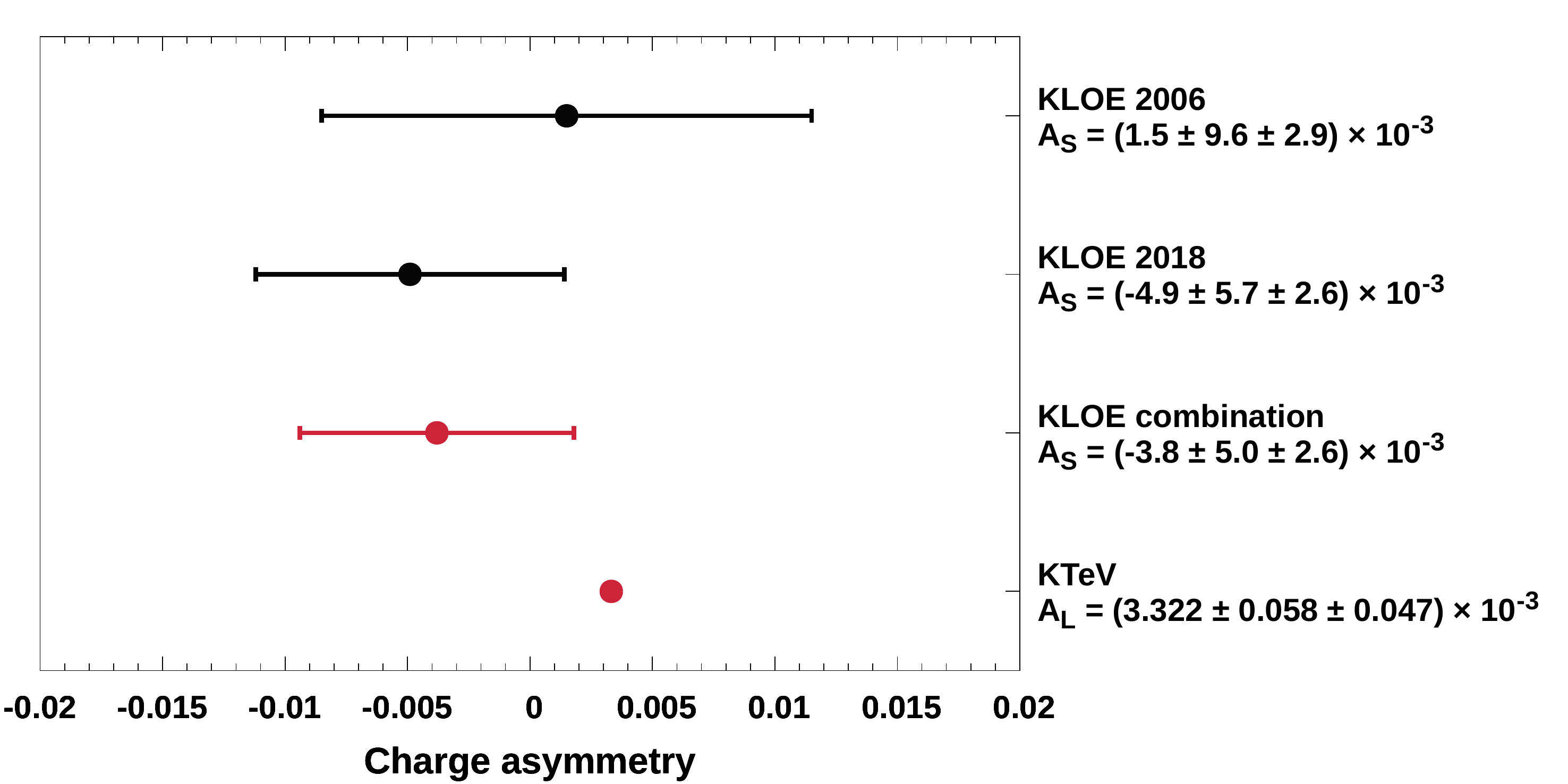}
      \caption{
Comparison of the previous 
result for $A_S$ 
(KLOE 2006 ~\cite{Ambrosino:2006si}), the result presented in this paper (KLOE 2018) and the combination of the two.
The KTeV result for $A_L$~\cite{AlaviHarati:2002bb} is also shown.
The uncertainties of the points correspond to the  statistical and systematic uncertainties summed in quadrature.
      }
      \label{fig::plotresultsas}
\end{figure}

Taking into account the correlations of the systematical uncertainties
of both measurements, based on similar analysis schemes, their combination provides:
\begin{equation}
    \asresultcombined~.
    \label{ascombined1}
\end{equation}
A comparison of these results is shown 
in Figure~\ref{fig::plotresultsas}.

The combined result \ref{ascombined1} together with the KTeV result on $A_L$~\cite{AlaviHarati:2002bb} 
yields for the sum and difference of asymmetries:
\begin{align}
    (A_S - A_L)/4 & = Re(\delta_K) + Re(x_-)= (-1.8 \pm 1.4 )\times 10^{-3},\\
    (A_S + A_L)/4 & = Re(\epsilon_K) -Re(y) = (-0.1 \pm 1.4) \times 10^{-3}.
\end{align}
Using $Re(\delta_K) = (2.5 \pm 2.3) \times 10^{-4}$~\cite{Patrignani:2016xqp}
and $Re(\epsilon_K) = (1.596 \pm 0.013) \times 10^{-3}$~\cite{Ambrosino:2006ek}
the $CPT$ violating parameters $Re(x_-)$ and $Re(y)$ are extracted: 
\begin{align}
    Re(x_-) & = (-2.0 \pm 1.4) \times 10^{-3}, \\
    Re(y) & = (1.7 \pm 1.4) \times 10^{-3},
\end{align}
which are consistent with $CPT$ invariance and improve by almost a factor of two the previous results~\cite{Ambrosino:2006si}. 

 \section*{Acknowledgments}
 We warmly thank our former KLOE colleagues for the access to the data collected during the KLOE data taking campaign.
 We thank the DA$\Phi$NE team for their efforts in maintaining low background running conditions and their collaboration during all data taking. We want to thank our technical staff: 
 G.F. Fortugno and F. Sborzacchi for their dedication in ensuring efficient operation of the KLOE computing facilities; 
 M. Anelli for his continuous attention to the gas system and detector safety; 
 A. Balla, M. Gatta, G. Corradi and G. Papalino for electronics maintenance; 
 C. Piscitelli for his help during major maintenance periods. 
 This work was supported in part 
 by the Polish National Science Centre through the Grants No.\
 2013/08/M/ST2/00323,
 2013/11/B/ST2/04245,
 2014/14/E/ST2/00262,
 2014/12/S/ST2/00459,
 2016/21/N/ST2/01727,
 2016/23/N/ST2/01293,
 2017/26/M/ST2/\-00697.

\bibliography{bibliography}

\end{document}